**Record magnetoresistance, enhanced superconductivity, and fermiology in $WTe_2$**

Gianluca Delgado,[1,†] Elliott Runburg,[1,†] Chaowei Hu,[1,†] Yuzhou Zhao,[1,2] Jonathan M. DeStefano,[1] Keng Tou Chu,[1] Florie Mesple,[1] Ellis Thompson,[1] Kenji Watanabe,[3] Takashi Taniguchi,[4] Jihui Yang,[2] Matthew Yankowitz,[1,2] Xiaodong Xu[1,2], Jiun-Haw Chu[1,*] and David Cobden[1,*]

[1]Department of Physics, University of Washington, Seattle, WA, 98195, USA
[2]Department of Materials Science and Engineering, University of Washington, Seattle, WA 98195 USA
[3]Research Center for Functional Materials, National Institute for Materials Science, 1-1 Namiki, Tsukuba 305-0044, Japan
[4]International Center for Materials Nanoarchitectonics, National Institute for Materials Science, 1-1 Namiki, Tsukuba 305-0044, Japan

[†]These authors contributed equally to this work.
[*]Corresponding authors: jhchu@uw.edu, dcobden@uw.edu

The diverse electronic properties of transition metal chalcogenides can be very sensitive to crystal imperfections. A new crystal growth technique, known as horizontal flux transport, offers a route to improved crystal quality. By refining this technique and applying it to the topological semimetal $WTe_2$, we achieved crystals with an order of magnitude less disorder as determined by electrical transport and scanning tunneling microscopy measurements. At low temperatures these crystals exhibit the largest magnetoresistance reported in a metal. Exfoliated monolayers show quantum oscillations for the first time in the electrostatically doped metallic states, enabling determination of band degeneracies and the valley splitting induced by an electric field. Moreover, they exhibit a gated superconducting dome with a greatly enhanced critical temperature approaching 1.8 K. This advance opens up new avenues for employing $WTe_2$ in topological electronics and gated superconducting devices, and promises comparable breakthroughs with other chalcogenides.

**Introduction.** Transition metal chalcogenides, in forms ranging from bulk crystals to exfoliated monolayer sheets, host diverse electronic phenomena such as charge density waves, superconductivity, diverse magnetic orders, Wigner crystals, topological features, and even fractional anomalous quantum Hall states. An excellent example is the semimetal $WTe_2$. In the bulk, it exhibits giant nonsaturating magnetoresistance[1,2] and its low symmetry, small carrier density, Weyl points, and strong spin-orbit coupling make it valued for spintronics[3–5] and electron hydrodynamics[6–8] studies. In the monolayer limit it exhibits quantum spin Hall edge modes,[9–13] gate-induced superconductivity,[14,15] intriguing thermoelectric anomalies[15,16] and a possible excitonic insulator ground state.[17,18] The sensitivity of these phenomena to disorder has motivated vigorous pursuit of improved chalcogenide crystal growth.[2,8,19–21] In particular, improvements to the chalcogen self-flux technique have reduced defect concentrations by orders of magnitude relative to conventional chemical vapor transport[2,19] allowing, for example, observation of the fractional quantum Hall effect in monolayer $WSe_2$.[22] Recently, a new technique known as horizontal flux transport (HFT)[23,24] has been introduced to overcome limitations of the self-flux technique.

In this work, we report dramatic consequences for the crystal quality and electronic properties of using HFT growth for $WTe_2$.[25–28] Our HFT-grown $WTe_2$ crystals have residual resistivity ratios (RRRs) of up to three times the previous record and magnetoresistance approaching a million-fold at 13 T, the largest ever reported for a metallic solid. The exfoliated monolayers are the first to exhibit quantum oscillations, confirming that both hole- and electron-doped metallic states are Fermi liquids and enabling determinations of degeneracies and measurement of the electric field-induced conduction band valley splitting. Moreover, the electron doping that induces the insulator-metal transition and superconductivity is reduced to a strikingly low value of $1.0 \times 10^{12}$ cm$^{-2}$, while the maximum superconducting critical temperature is increased from below[14,15] 1 K to nearly 1.8 K, making it accessible for the first time at helium-4 temperatures. The findings show that disorder plays a controlling role in the electronic response of $WTe_2$ and imply that HFT growth could enable major advances with other chalcogenides.

The key differences between conventional self-flux and HFT growth are indicated on the temperature-composition phase diagram shown in Fig. 1a. In the former, a saturated solution of W in Te is gradually cooled from $T_i$ to $T_f$ and crystals grow while the flux composition tracks the solubility boundary (green line). In the latter[23] a temperature gradient is applied to a tilted sealed ampule with a neck in it, as shown in Fig. 1b. The source is trapped on the hotter side of the neck where it gradually dissolves at temperature $T_1$ (black star), and the W diffuses in the flux through the neck to the cooler side where the crystals grow at temperature $T_2$ (black circle). HFT growth can greatly increase the yield, which is limited only by the amount of starting metal. However, it has important additional benefits in that the crystals grow at a constant temperature and flux composition, and the operating temperatures are lower, all of which can reduce defect and impurity concentrations. More details are given in Methods.

**Bulk crystals.** We characterized 30 HFT-grown $WTe_2$ crystals from four batches, grown using Te and W sources of varying purity and with different temperature parameters (see Methods). Figure 1c shows the temperature dependence of the zero-field four-terminal resistivity, $\rho_0$, for a crystal exhibiting an RRR of 12200. Figure 1d shows the dependence of the resistivity $\rho$ on magnetic field $B$ at temperature $T$ =100 mK. The ratio $\rho/\rho_0$ reaches $7.8 \times 10^5$ at $B = 13$ T, exceeding that for any other metallic solid ever reported.[29,30] The magnetoresistance, $MR = (\rho - \rho_0)/\rho_0$, is approximately quadratic over the full range (see inset log-log plot), and fitting it to $\mu^2 B^2$ (red dashed line) yields a value for the mobility parameter $\mu$ of 70 T$^{-1}$= $7 \times 10^5$ cm$^2$V$^{-1}$s$^{-1}$. On the other hand, estimating the quantum mobility from the lowest magnetic field where Shubnikov-de Haas (SdH) oscillations are visible (about 3 T) gives only ~0.3 T$^{-1}$ = 3000 cm$^2$V$^{-1}$s$^{-1}$, similar to that for flux-grown crystals[1]. Such a large difference between quantum mobility (small-angle scattering) and transport mobility (momentum relaxation) is not unprecedented in semimetals (see Methods).[31] Notably, the quantum mobility remains nearly constant across crystals spanning a wide range of transport mobilities (SI7), implying that the two are limited by distinct scattering mechanisms. Figure 1e displays measurements of $\mu$ and the RRR for all the crystals. Most of them exceed the previous record (black square),[2,8] but the scatter suggests that additional factors remain to be identified, such as the effects of different defect species. Figure 1f shows a scanning tunneling topography image of a cleaved surface of an HFT crystal in which just two surface defects are seen in a (90 nm)$^2$ area. From multiple images we estimated the average defect density in this crystal to be $(2.9 \pm 0.2) \times 10^{-4}$ per unit cell (or $1.5 \times 10^{11}$ cm$^{-2}$ per layer; see SI1).

**Monolayers.** Monolayers exfoliated from HFT crystals were incorporated into devices with the structure indicated in the upper inset of Fig. 2a. It is challenging to prevent them degrading by oxidation, and all the data shown here were taken from our best device (see SI2). We first present measurements made in magnetic fields sufficient to fully suppress superconductivity (which will be discussed below) and remnant quantum-spin Hall edge conduction. The main part of Fig. 2a is a map of the four-terminal resistance $R_{xx}$ vs net electron doping $n_e$ and displacement field $D_\perp$ (see SI2 for definitions) taken at $T \approx 100$ mK and $B = 9$ T. The sample is insulating (dark red) near $n_e = 0$ but becomes metallic above a doping threshold at $n_e = n_c$ on the electron side and $n_e = -p_c$ on the hole side. This matches the usual behavior of monolayer $WTe_2$, except in that both $n_c$ and $p_c$ are smaller than in any prior reports. (Note that $p_c$ is affected by the contacts going insulating, and this explains its variation with $D_\perp$). Importantly, unlike in any previous samples, oscillations periodic in $n_e$ can be seen in the metallic regions. These SdH oscillations appear at ~3 T, indicating a quantum mobility similar to that of the bulk crystals discussed above. In contrast with the bulk, we estimate this to be comparable with the transport mobility (allowing for uncertainty in sample geometry), indicating a mean free path on the order of 1 μm (see SI3). The primary oscillation period in $n_e$ is $2eB/h$ on the hole side and $4eB/h$ on the electron side, as indicated by combs of vertical lines, thus confirming the expected double spin degeneracy of the valence band edge at $\Gamma$ and the four-fold degeneracy of the conduction band (CB) edge due to the valleys at $\Lambda$ and $\Lambda'$ (see lower inset in Fig. 2a).

At $D_\perp = 0$, the magnetic field dependence on the electron-doped side, shown in Fig. 2b, displays a single dominant Landau fan (dashed lines). The straightness of the fan shows that $n_e$ is linear in both gate voltages, and thereby provides the calibration for the $n_e$ axis that we have used throughout. Indications of a halving of the period at higher field (dotted lines) are consistent with spin-degeneracy lifting, as indicated in the inset diagram. Additionally, in Fig. 2a we notice that on the electron-doped side the oscillation amplitude varies with $D_\perp$, passing through a minimum near $D_\perp = -0.7$ V/nm. Figure 2c shows the magnetic field dependence at $D_\perp = -1.25$ V/nm, which can be matched with a superposition of two copies of the same fan (indicated using black and red dotted lines) displaced in density by $\Delta n_e \approx 5 \times 10^{11}$ cm$^{-2}$ (see SI4). This can be explained by a Rashba-like effect of the electric field, which breaks the centrosymmetry of the monolayer and linearly splits the valleys for each spin (or, alternatively, the spins for each valley), as indicated in the inset to Fig. 2c. At $D_\perp = -0.7$ V/nm the oscillations from the split branches are out of phase. Combining $\Delta n_e$ with our previously measured total CB density of states,[18] $N_e = 3.7 \times 10^{11}$ cm$^{-2}$meV$^{-1}$, we can estimate the valley splitting to be $\Delta E_v = \Delta n_e / N_e \approx 1.4$ meV. Note that such splitting should be absent[32] near $\Gamma$, and, indeed, no amplitude variation with $D_\perp$ is seen on the hole side in Fig. 2a.

Figure 3a is a map of $R_{xx}$ like that in Fig. 2a but at $B = 0$. Now there is a dark blue region on the right where $R_{xx}$ is very small, signaling superconductivity. When $R_{xx}$ is measured vs $n_e$ as the temperature is lowered (Fig. 3b), a suppression is first noticed at 1.7 K for $n_e \sim 5 \times 10^{12}$ cm$^{-2}$, which we call $n_m$. At lower $T$, $R_{xx}$ is below the noise floor whenever $n_e$ is above a threshold which decreases to $\sim 1.0 \times 10^{12}$ cm$^{-2}$ at the nominal base temperature of 20 mK. The critical temperature, $T_c$, defined by where the resistance is half that of the normal state, is plotted against $n_e$ in Fig. 3c (black circles). It rises rapidly from zero above a threshold of $n_c = 1.0 \times 10^{12}$ cm$^{-2}$, reaches a maximum of $\sim 1.75$ K at $n_m$, then gradually decreases at higher doping tracing out an asymmetric dome. On the same graph we also plot measurements of $T_c$ from other devices in the literature.[14,15,33] The large variation between them implies a strong sensitivity to disorder and has a systematic appearance: smaller $n_c$ is associated with higher $T_c$ at all densities, and a dome is only seen in the two cases when $n_c$ is below $n_m$.

The normal state is restored by applying either a magnetic field $B$ or a current bias $I_{bias}$. Figure 3d shows the variation with $n_e$ of the critical field $B_c$, defined by where the resistance is half that of the normal state (see left inset). At 100 mK, $B_c$ rises sharply to a maximum of ~200 mT at a doping smaller than $n_m$ before falling off steadily. At 1.4 K, $B_c$ is an order of magnitude smaller and peaked near $n_m$. The corresponding Landau-Ginzburg coherence length, $\xi_{GL} = \sqrt{\phi_0 / 2\pi B_c(T \to 0)}$, where $\phi_0 = \hbar/2e$, is plotted vs $n_e$ in the right inset (see SI5 for details). At larger $n_e$ (above $n_{max}$) it is an order of magnitude larger than the electron spacing $n_e^{-1/2}$ (black dashed line), motivating a comparison with BCS theory. Indeed, the BCS result in the clean limit, $\xi_{BCS} = \hbar v_F / \pi\Delta = 0.41\sqrt{n_e}/(g N_e k_B T_c)$, with $v_F$ the Fermi velocity, $\Delta$ the zero-temperature gap, $k_B$ Boltzmann's constant, and conduction band degeneracy $g = 4$, matches $\xi_{GL}$ as function of $n_e$ fairly well up to a factor of 2 (red dotted line). This suggests that the clean limit holds, consistent with $\xi_{BCS}$ being shorter than the mean free path of ~1 $\mu$m estimated above.

Figure 3e is a map of differential 4-terminal resistance vs dc bias current $I_{bias}$ and $n_e$ at 100 mK. There is variation between different sets of contacts (see SI6), but in all cases the critical current $I_c$ at which resistance suddenly appears is symmetric about zero current and its increase with $n_e$ is superlinear near $n_c$ but linear or sublinear at higher doping. For several choices of contacts including the one shown here, we notice that at higher doping $I_c$ becomes proportional to $n_e$ (red dashed line), and not to $n_e - n_c$. Although the mechanism that determines $I_c$ is uncertain, this hints that the electrons that are localized in the insulating state (when $n_e < n_c$) participate in the superconducting state (when $n_e > n_c$).

**Discussion.** The above measurements show that HFT crystal growth has a dramatic impact on the electronic properties of $WTe_2$. In the bulk, which is of interest for spintronics, electron hydrodynamics and Weyl physics, the magnetoresistance retains the typical nonsaturating quadratic behavior but is enhanced to a level unprecedented in a metal. In the monolayer limit, the metallic state now shows SdH oscillations, enabling fermiology and measurement of the electric field-induced valley splitting, which is accompanied by induced Berry curvature.[32] The lowering of the doping thresholds $n_c$ and $p_c$ shows that disorder plays a role in the insulating state, although both remain considerably larger than the STM-visible defect density which amounts to ~$3 \times 10^{-4}$ per 22 Å$^2$ unit cell ≈ $1.5 \times 10^{11}$ cm$^{-2}$ in a monolayer. This could be related to excitonic correlations[17,18] combining with weak disorder to localize carriers. We note that the conductance of the edge states was no higher than in earlier flux-grown samples, being undetectable at 100 mK, supporting the idea that it is dominated by the inevitable disorder of the torn edge.

Perhaps the most impactful finding is the enhancement of the gate-induced monolayer superconductivity, which now persists to helium-4 temperatures and higher fields and currents. While at larger $n_e$ it appears BCS-like, the sensitivity to disorder could indicate non-s-wave pairing. It also occurs at a remarkably small Fermi energy $E_F{\sim}n_c/N_e \approx 2.5$ meV, implying a low superfluid density and hence high stiffness favorable to observing a Berezinski-Kosterlitz-Thouless transition. It also implies violation of the retardation requirement for phonon-mediated pairing and favors electronic intermediaries such as excitons.[34,35] The reduced disorder should benefit studies of the quantum phase transition between these contrasting states[15] and may resolve questions concerning a postulated neutral-particle Fermi surface in the insulating state.[16] The more robust superconductivity also presents new possibilities for gated superconducting junctions and circuits, being highly tunable and readily coupled to other 2D electronic systems such as quantum spin Hall and fractional quantum anomalous Hall states.[24]

## Methods

**Horizontal flux growth.** An ampule with a narrow neck part way along (see Fig. 1b) is sealed under vacuum with the high purity starting materials, W and Te, positioned on one side of the neck and placed in a two-zone furnace. The ampule is first heated to a temperature of 500 °C at which the Te flux melts and flows to form two ponds connected through the neck. The W quickly reacts to form a polycrystalline telluride. To initiate growth, the furnace is tilted by ~10 degrees and a temperature gradient is applied such that the lower pond is cooled to the growth temperature $T_2$ (~550 °C). The source solid is trapped by the neck in the upper pond at a higher temperature $T_1$ (~600 °C) where it gradually dissolves. In the cooler lower pond, the solubility is lower and $WTe_2$ crystals nucleate and grow at a rate limited by diffusion of W through the neck from the source. This is reminiscent of vapor transport growth employing an evaporated source, carrier gas, and a temperature differential between source and growth regions. To terminate growth and extract the crystals from the flux, the tilt of the tube is carefully reversed so that all the flux flows to the left side, leaving the $WTe_2$ crystals trapped on the right side of the neck. A HFT growth typically takes one month to complete.

In standard flux growth, the temperature of a solution of W dissolved in Te is gradually decreased from $T_i$, causing the solubility to decrease. When it becomes saturated, crystals nucleate and subsequently the metal concentration in the flux changes tracking the solubility line, as indicated by the green line drawn on the binary phase diagram[36] in Fig. 1a. This procedure has several inherent problems. First, since different regions of a crystal grow under different conditions, the defect distribution is not uniform. Second, the low solubility of the metal requires use of the highest possible initial temperature compatible with the tolerable vapor pressure of the chalcogen (Te reaches 1 atm at ~1000 °C), limiting the yield, increasing defect populations, and promoting contamination by container wall material (typically silica).[6] Third, the existence of other binary phases necessitates avoiding certain temperature windows.

By contrast, in HFT growth, the temperature $T_2$, and the composition of the solution are constant, as indicated by the black dot on the phase diagram. This incurs advantages similar to those offered by the Czochralski method used for silicon. First, the kinetics are constant throughout the growth and the defect density in the crystals is thus uniform. Second, the yield is limited only by the amount of starting metal. Third, both $T_1$ and $T_2$ can be kept low enough to avoid dangerous chalcogen vapor pressure, reduce contamination and defect density, avoid the formation of other binary phases, and to control the growth rate. Figure 1e shows measurements on four (growth 1 [red dots], growth 2 [black diamonds], and growth 3 [blue squares]. In each case, 30-40 g of tellurium (Sigma-Aldrich, 99.9999%) was combined with tungsten foil (Sigma-Aldrich, 99.95+%) /powder (Sigma-Aldrich, 99.99+%) at a molar ratio of 20:1. The tungsten foil (growths #1 and 2) or pellet pressed from the powder (growth #3, 4) was pre-annealed in an Ar + 5% $H_2$ environment overnight at 1000 °C before growth. For the four batches, $T_1$ and $T_2$ were, respectively, 600 and 580°C; 700 and 690°C; 600 and 580°C; 560 and 530°C.

Last, the simple flux extraction process (Fig. 1b) avoids the need for a quartz wool filter or a complex crucible design[6]. In cases where we need to remove residual flux, we reseal the extracted crystals near one end of a separate quartz tube and heat this end to ~300°C overnight to evaporate the flux while keeping the other end at room temperature to condense it.

**Single crystal measurements.** As-grown $WTe_2$ crystals were made into a standard 4-terminal, or 6-terminal resistance bar using gold wire and silver paste. The resistance was measured using pre-amplifier and lock-in amplifiers at 97 Hz with varying current from $10\mu A$ to 1mA to accommodate the large resistance change in Quantum design physical property measurement system (QD PPMS) and Bluefors, or simply using built-in resistance module in QD PPMS. A detailed analysis of the magnetotransport using a two band model[37] is presented in SI7.

**Scanning tunneling microscopy.** STM measurements were performed in a Scienta Omicron Polar system operating at around 5 K. Bulk $WTe_2$ crystals were cleaved in ultra high vacuum directly before the measurements. Chemically etched tungsten tips were calibrated against a single crystal Cu(111) surface in advance and sometimes resharpened on the $WTe_2$ via voltage pulses as needed. During measurements, a voltage equal to -$V_b$ with respect to the grounded sample was applied to the tip, and tunneling current $I$ was collected downstream from the tip. The measured topography images were FFT filtered when needed to remove very low-frequency components associated with vibrational noise. The defect count was collected across multiple macroscopic areas in each of multiple cleaves (see SI1).

**Monolayer device fabrication and measurements.** The device was assembled in an argon glovebox with $O_2$ and $H_2O$ levels <0.1 ppm. Monolayer $WTe_2$ was obtained by exfoliating bulk crystals of HFT-grown $WTe_2$ onto 285 nm $SiO_2$ and identifying flakes through optical contrast under a microscope. Other devices were fabricated too, but due to variations in glove box condition and preparation time their quality was poor. A polycarbonate stamp was then used to pick up the top gate graphite, four nanometers thick hBN, and the $WTe_2$ monolayer in that order before depositing onto pre-patterned platinum electrodes atop a pre-assembled graphite-hBN back gate. The mixing chamber base temperature of the Bluefors LD250 dilution refrigerator was 10 mK, though the device electron temperature at base was probably closer to 100 mK (see Fig. 3b and SI3.) The four-terminal resistance was measured using lock-in amplifiers at 18 Hz and a constant-current configuration, keeping all signals small enough for linear response at 100 mK. Gate voltages were applied using Keithley 2450 voltage sources.

### Author Contributions

LD and ER fabricated monolayer $WTe_2$ FET samples. LD, ER and CH performed the transport measurements and analysis. CH, YZ and JMD grew and characterized the crystals. KTC, FM and ET performed STM and analysis. KW and TT provided hBN crystals. MY, XX, JHC, JY and DC conceived and directed the project. LD and DC wrote the paper with contributions from all.

### Acknowledgements

Research on HFT $WTe_2$ crystals and their properties at UW is supported as part of Programmable Quantum Materials, an Energy Frontier Research Center funded by the US Department of Energy (DOE), Office of Science, Basic Energy Sciences (BES), under award DE-SC0019443. LD was partially supported by the U.S. National Science Foundation through the UW Molecular Engineering Materials Center (MEM-C), a Materials Research Science and Engineering Center (DMR-1719797 and DMR-2308979). Device fabrication made use of facilities and instrumentation provide by the MEM-C Shared Facilities. We thank P. Nguyen and M. Rontani for discussions and Kyle Cobden for graphics assistance.

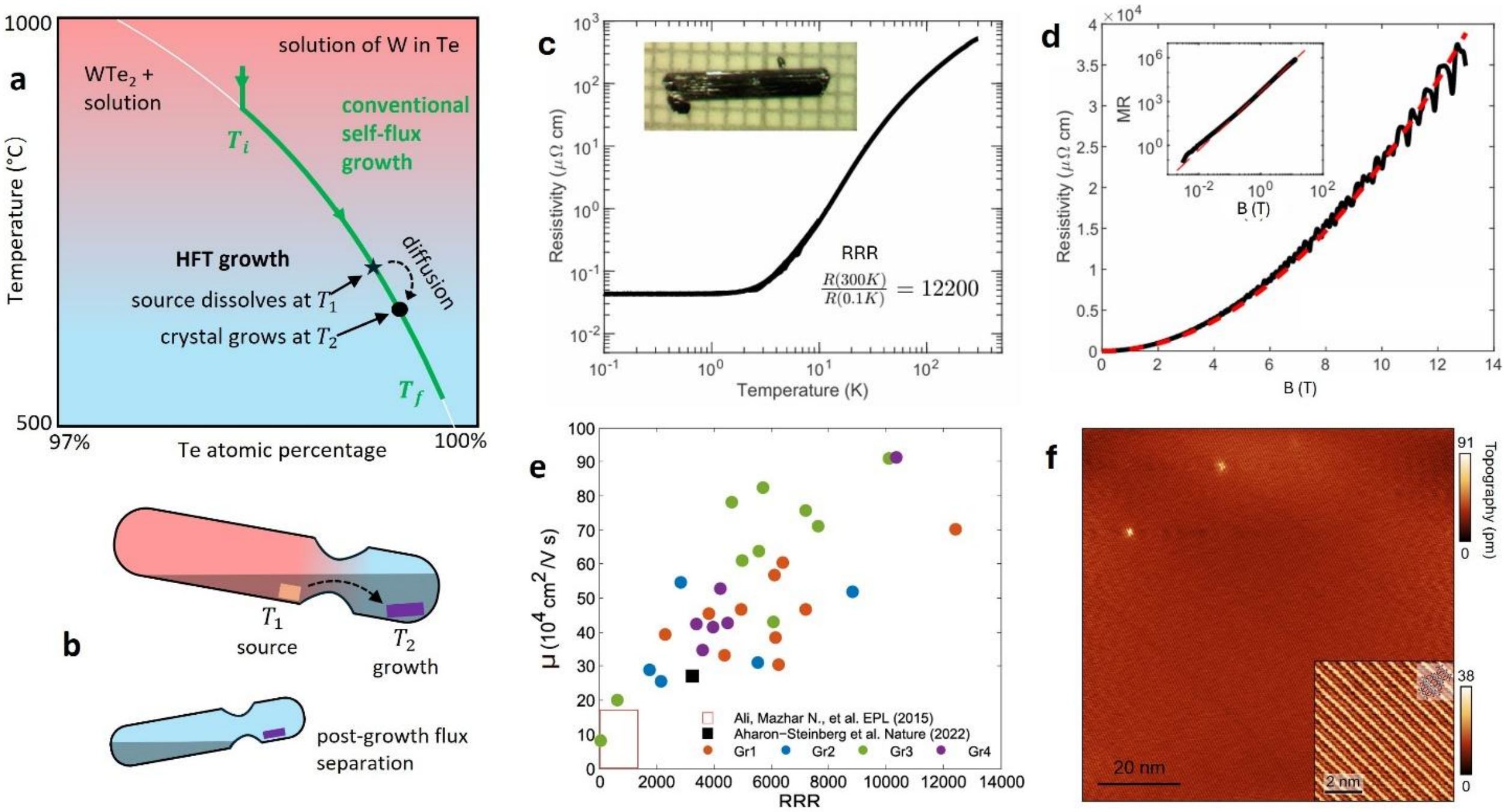

**Figure 1. Horizontal flux transfer (HFT) growth and characterization of $WTe_2$ crystals. a.** Binary phase diagram of $WTe_2$ contrasting growth techniques. In conventional self-flux growth, a starting solution of W in Te is steadily cooled from temperature $T_i$ to $T_f$ and the crystals form as the saturated solution tracks the solubility line (green). In HFT growth, the source dissolves at $T_1$ (black star) and crystals grow at a constant temperature $T_2$ and flux composition (black dot). **b.** (Top) sketch of how this is achieved, by applying a temperature gradient to a tilted sealed ampule with a neck which traps the source metal at higher temperature $T_1$, where it gradually dissolves, while crystals grow below the neck at temperature $T_2$ as W diffuses over from the source (dashed arrow). (Bottom) the crystals can be separated from the flux by reversing the tilt. **c.** Temperature dependence of the zero-field resistivity $\rho_0(T)$ of a crystal (inset: photograph with 1 mm grid) with a residual resistivity ratio (RRR) of 12200. **d.** Resistivity $\rho$ vs magnetic field $B$ for this crystal at 0.1 K, in linear scale. Inset: log-log plot of the magnetoresistance, $MR = (\rho - \rho_0)/\rho_0$, at 100 mK, with $\rho_0 = 0.040$ μΩcm. The red dashed lines in both main graph and inset corresponds to the best fit to $MR = \mu^2 B^2$ with fitting parameter $\mu = 70$ $m^2V^{-1}s^{-1}$. **e.** $\mu$ plotted vs RRR for 30 crystals from four different batches (indicated by color) of HFT-grown $WTe_2$. The red box includes all the values reported in Ref. 2 and the black square shows the record from Ref. 8. **f.** STM image of a cleaved crystal, showing two surface-layer defects in a 90×90 $nm^2$ area. Inset: higher resolution image with lattice identification.

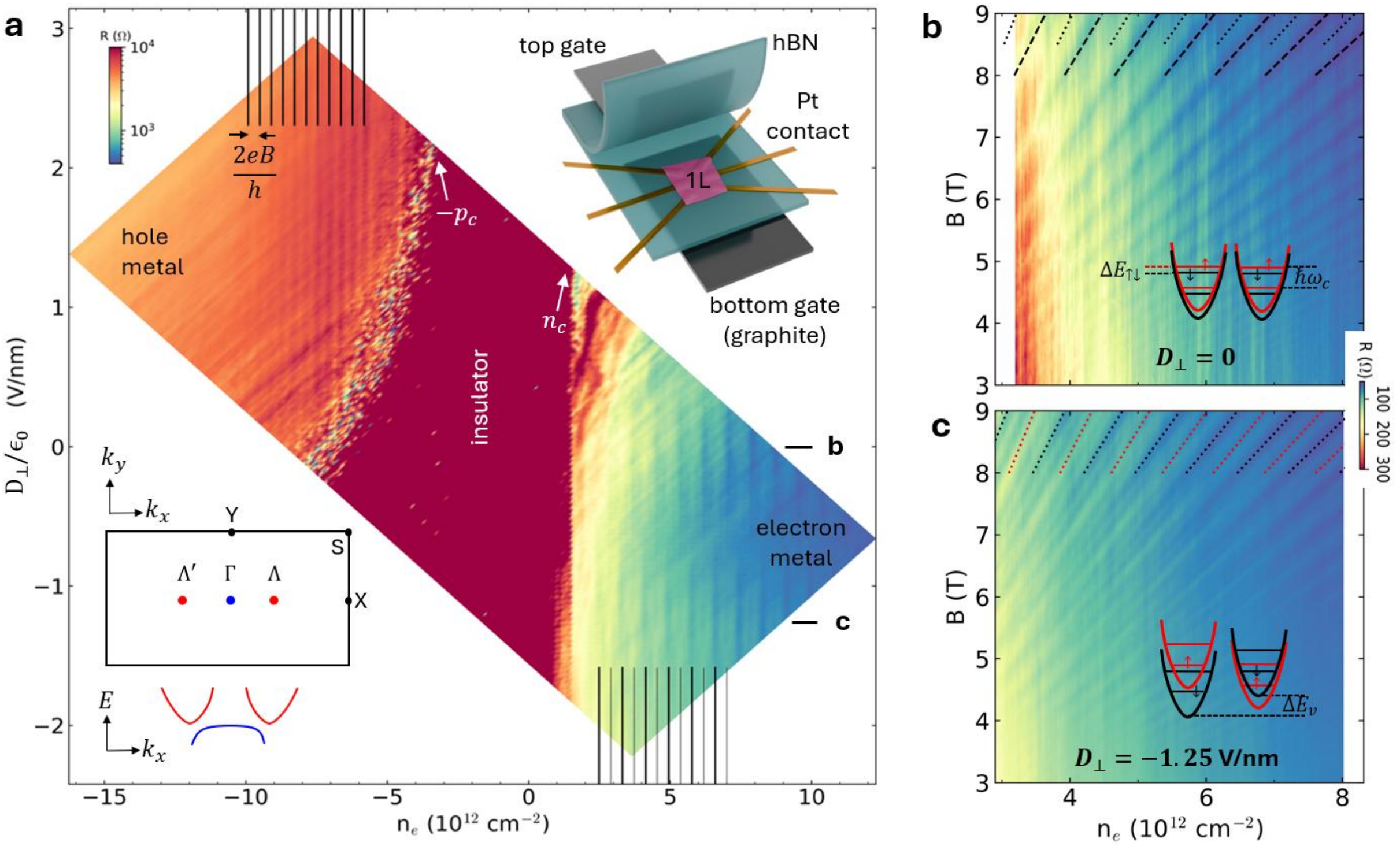

**Figure 2. Quantum oscillations in monolayer WTe₂ at high magnetic fields. a.** 4-terminal longitudinal resistance $R_{xx}$ vs gate-doping $n_e$ and perpendicular displacement field $D_\perp$ at $B = 9$ T for device 1. The combs are drawn to show how the fundamental period of the SdH oscillations (spacing of the thicker lines) on the electron-doped side (right) is twice that on the hole-doped side (left). Upper inset: cartoon of monolayer device structure. Lower inset: Brillouin zone and sketch of bands along the zone axis at $B = 0$. **b.** $R_{xx}$ vs $n_e$ and $B$ at $D_\perp = 0$. A single Landau fan dominates, with minima indicated by the dashed lines, and evidence for spin degeneracy lifting is visible at higher fields (dotted lines), as indicated in the inset sketch. **c.** Similar map taken at $D_\perp = -1.25$ V/nm. What is seen here can be interpreted as two superposed Landau fans (red and black dotted lines) offset in density as a result of valley splitting $\Delta E_v$, the situation indicated in the inset sketch.

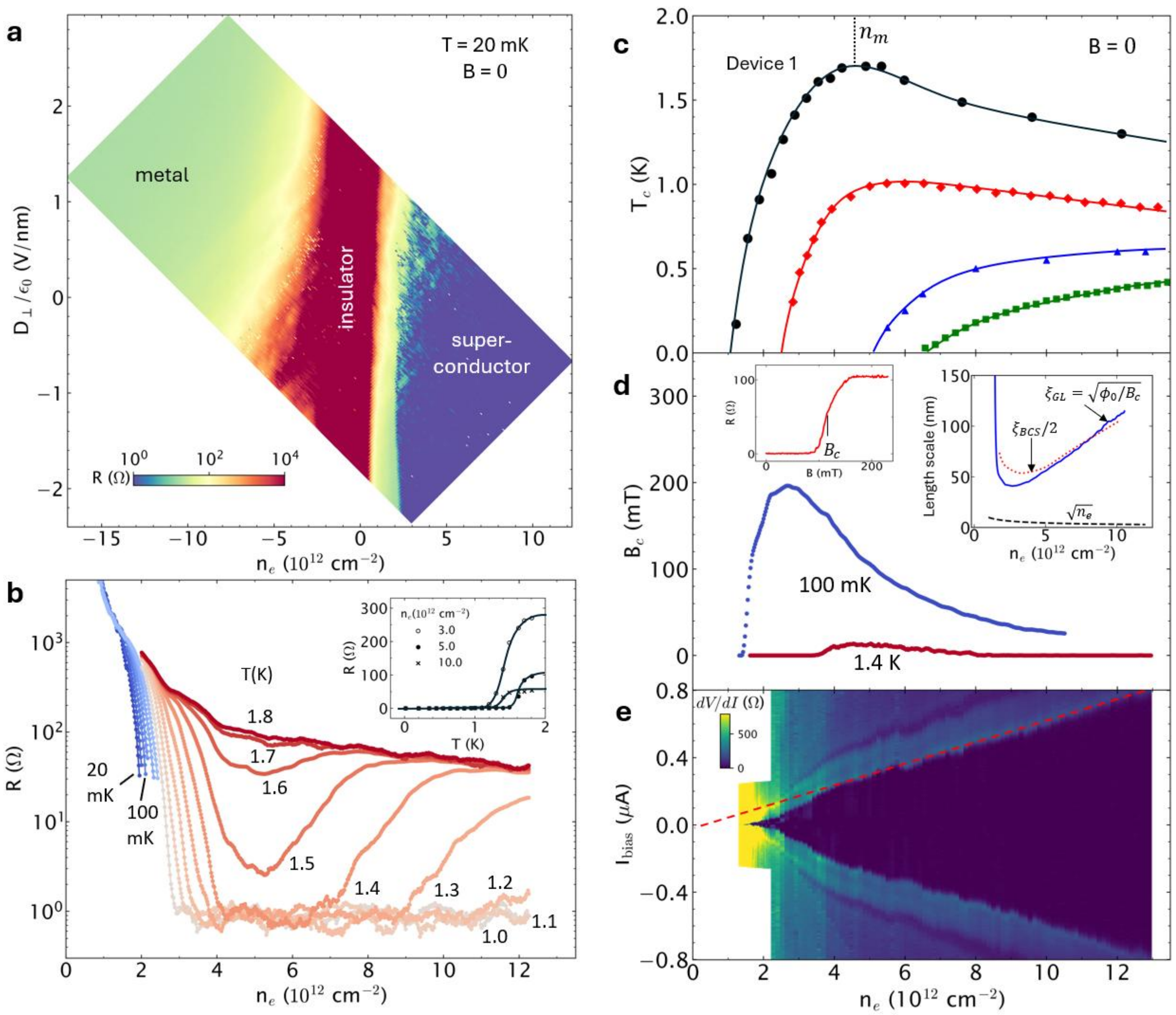

**Figure 3. Enhanced gated superconductivity. a.** $R_{xx}$ vs gate doping $n_e$ and displacement field $D_\perp$ at $B = 0$ and base temperature. **b.** $R_{xx}$ vs $n_e$ at $D_\perp = 0$ for the indicated temperatures. Inset: $R_{xx}$ vs $T$ at the indicated dopings. Solid lines are guides for the eye. **c.** Critical temperature $T_c$ vs $n_e$ (black circles) for Device 1, and corresponding measurements from three monolayer $WTe_2$ devices reported in the literature (red diamonds [Ref. 33], blue triangles [Ref. 14], green squares [Ref. 15]). All solid lines here are guides for the eye. **d.** Variation of critical field $B_c$ with $n_e$ at the indicated temperatures. Left inset: a trace of $R_{xx}$ vs $B$ at 100 mK and $n_e = 4.6 \times 10^{12}$ cm$^{-2}$, showing $B_c$ defined at half the normal state resistance. Right inset: comparison of coherence length derived from $B_c$ ($\xi_{GL}$) with that derived from $T_c$ ($\xi_{BCS}$) and with the interelectron spacing ($n_e^{1/2}$). **e**. Two-terminal differential resistance vs current bias and doping for a pair of adjacent contacts. At higher doping the critical current becomes proportional to $n_e$ (red dashed line).

## References

(1) Ali, M. N.; Xiong, J.; Flynn, S.; Tao, J.; Gibson, Q. D.; Schoop, L. M.; Liang, T.; Haldolaarachchige, N.; Hirschberger, M.; Ong, N. P.; Cava, R. J. Large, Non-Saturating Magnetoresistance in $WTe_2$. *Nature* **2014**, *514* (7521), 205–208. https://doi.org/10.1038/nature13763.

(2) Ali, M. N.; Schoop, L.; Xiong, J.; Flynn, S.; Gibson, Q.; Hirschberger, M.; Ong, N. P.; Cava, R. J. Correlation of Crystal Quality and Extreme Magnetoresistance of $WTe_2$. *Europhys. Lett.* **2015**, *110* (6), 67002. https://doi.org/10.1209/0295-5075/110/67002.

(3) Rhodes, D.; Das, S.; Zhang, Q. R.; Zeng, B.; Pradhan, N. R.; Kikugawa, N.; Manousakis, E.; Balicas, L. Role of Spin-Orbit Coupling and Evolution of the Electronic Structure of $WTe_2$ under an External Magnetic Field. *Phys. Rev. B* **2015**, *92* (12), 125152. https://doi.org/10.1103/PhysRevB.92.125152.

(4) MacNeill, D.; Stiehl, G. M.; Guimaraes, M. H. D.; Buhrman, R. A.; Park, J.; Ralph, D. C. Control of Spin–Orbit Torques through Crystal Symmetry in $WTeWTe_2$/Ferromagnet Bilayers. *Nat. Phys.* **2017**, *13* (3), 300–305. https://doi.org/10.1038/nphys3933.

(5) Peng, C.-W.; Liao, W.-B.; Chen, T.-Y.; Pai, C.-F. Efficient Spin-Orbit Torque Generation in Semiconducting WTe2 with Hopping Transport. *ACS Appl. Mater. Interfaces* **2021**, *13* (13), 15950–15957. https://doi.org/10.1021/acsami.1c03530.

(6) Vool, U.; Hamo, A.; Varnavides, G.; Wang, Y.; Zhou, T. X.; Kumar, N.; Dovzhenko, Y.; Qiu, Z.; Garcia, C. A. C.; Pierce, A. T.; Gooth, J.; Anikeeva, P.; Felser, C.; Narang, P.; Yacoby, A. Imaging Phonon-Mediated Hydrodynamic Flow in $WTe_2$. *Nat. Phys.* **2021**, *17* (11), 1216–1220. https://doi.org/10.1038/s41567-021-01341-w.

(7) Xie, W.; Yang, F.; Xu, L.; Li, X.; Zhu, Z.; Behnia, K. Purity-Dependent Lorenz Number, Electron Hydrodynamics and Electron-Phonon Coupling in WTe2. *Sci. China Phys. Mech. Astron.* **2024**, *67* (8), 287014. https://doi.org/10.1007/s11433-024-2404-0.

(8) Aharon-Steinberg, A.; Völkl, T.; Kaplan, A.; Pariari, A. K.; Roy, I.; Holder, T.; Wolf, Y.; Meltzer, A. Y.; Myasoedov, Y.; Huber, M. E.; Yan, B.; Falkovich, G.; Levitov, L. S.; Hücker, M.; Zeldov, E. Direct Observation of Vortices in an Electron Fluid. *Nature* **2022**, *607* (7917), 74–80. https://doi.org/10.1038/s41586-022-04794-y.

(9) Qian, X.; Liu, J.; Fu, L.; Li, J. Quantum Spin Hall Effect in Two-Dimensional Transition Metal Dichalcogenides. *Science* **2014**, *346* (6215), 1344–1347. https://doi.org/10.1126/science.1256815.

(10) Tang, S.; Zhang, C.; Wong, D.; Pedramrazi, Z.; Tsai, H.-Z.; Jia, C.; Moritz, B.; Claassen, M.; Ryu, H.; Kahn, S.; Jiang, J.; Yan, H.; Hashimoto, M.; Lu, D.; Moore, R. G.; Hwang, C.-C.; Hwang, C.; Hussain, Z.; Chen, Y.; Ugeda, M. M.; Liu, Z.; Xie, X.; Devereaux, T. P.; Crommie, M. F.; Mo, S.-K.; Shen, Z.-X. Quantum Spin Hall State in Monolayer 1T'-$WTe_2$. *Nat. Phys.* **2017**, *13* (7), 683–687. https://doi.org/10.1038/nphys4174.

(11) Fei, Z.; Palomaki, T.; Wu, S.; Zhao, W.; Cai, X.; Sun, B.; Nguyen, P.; Finney, J.; Xu, X.; Cobden, D. H. Edge Conduction in Monolayer $WTe_2$. *Nat. Phys.* **2017**, *13* (7), 677–682. https://doi.org/10.1038/nphys4091.

(12) Wu, S.; Fatemi, V.; Gibson, Q. D.; Watanabe, K.; Taniguchi, T.; Cava, R. J.; Jarillo-Herrero, P. Observation of the Quantum Spin Hall Effect up to 100 Kelvin in a Monolayer Crystal. *Science* **2018**, *359* (6371), 76–79. https://doi.org/10.1126/science.aan6003.

(13) Shi, Y.; Kahn, J.; Niu, B.; Fei, Z.; Sun, B.; Cai, X.; Francisco, B. A.; Wu, D.; Shen, Z.-X.; Xu, X.; Cobden, D. H.; Cui, Y.-T. Imaging Quantum Spin Hall Edges in Monolayer $WTe_2$. *Sci. Adv.* **2019**, *5* (2), eaat8799. https://doi.org/10.1126/sciadv.aat8799.

(14) Sajadi, E.; Palomaki, T.; Fei, Z.; Zhao, W.; Bement, P.; Olsen, C.; Luescher, S.; Xu, X.; Folk, J. A.; Cobden, D. H. Gate-Induced Superconductivity in a Monolayer Topological Insulator. *Science* **2018**, *362* (6417), 922–925. https://doi.org/10.1126/science.aar4426.

(15) Song, T.; Jia, Y.; Yu, G.; Tang, Y.; Wang, P.; Singha, R.; Gui, X.; Uzan-Narovlansky, A. J.; Onyszczak, M.; Watanabe, K.; Taniguchi, T.; Cava, R. J.; Schoop, L. M.; Ong, N. P.; Wu, S.

Unconventional Superconducting Quantum Criticality in Monolayer $WTe_2$. *Nat. Phys.* **2024**, *20* (2), 269–274. https://doi.org/10.1038/s41567-023-02291-1.
(16) Wu, S.; Schoop, L. M.; Sodemann, I.; Moessner, R.; Cava, R. J.; Ong, N. P. Charge-Neutral Electronic Excitations in Quantum Insulators. *Nature* **2024**, *635* (8038), 301–310. https://doi.org/10.1038/s41586-024-08091-8.
(17) Jia, Y.; Wang, P.; Chiu, C.-L.; Song, Z.; Yu, G.; Jäck, B.; Lei, S.; Klemenz, S.; Cevallos, F. A.; Onyszczak, M.; Fishchenko, N.; Liu, X.; Farahi, G.; Xie, F.; Xu, Y.; Watanabe, K.; Taniguchi, T.; Bernevig, B. A.; Cava, R. J.; Schoop, L. M.; Yazdani, A.; Wu, S. Evidence for a Monolayer Excitonic Insulator. *Nat. Phys.* **2022**, *18* (1), 87–93. https://doi.org/10.1038/s41567-021-01422-w.
(18) Sun, B.; Zhao, W.; Palomaki, T.; Fei, Z.; Runburg, E.; Malinowski, P.; Huang, X.; Cenker, J.; Cui, Y.-T.; Chu, J.-H.; Xu, X.; Ataei, S. S.; Varsano, D.; Palummo, M.; Molinari, E.; Rontani, M.; Cobden, D. H. Evidence for Equilibrium Exciton Condensation in Monolayer $WTe_2$. *Nat. Phys.* **2022**, *18* (1), 94–99. https://doi.org/10.1038/s41567-021-01427-5.
(19) Edelberg, D.; Rhodes, D.; Kerelsky, A.; Kim, B.; Wang, J.; Zangiabadi, A.; Kim, C.; Abhinandan, A.; Ardelean, J.; Scully, M.; Scullion, D.; Embon, L.; Zu, R.; Santos, E. J. G.; Balicas, L.; Marianetti, C.; Barmak, K.; Zhu, X.; Hone, J.; Pasupathy, A. N. Approaching the Intrinsic Limit in Transition Metal Diselenides via Point Defect Control. *Nano Lett.* **2019**, *19* (7), 4371–4379. https://doi.org/10.1021/acs.nanolett.9b00985.
(20) Liu, S.; Liu, Y.; Holtzman, L.; Li, B.; Holbrook, M.; Pack, J.; Taniguchi, T.; Watanabe, K.; Dean, C. R.; Pasupathy, A. N.; Barmak, K.; Rhodes, D. A.; Hone, J. Two-Step Flux Synthesis of Ultrapure Transition-Metal Dichalcogenides. *ACS Nano* **2023**, *17* (17), 16587–16596. https://doi.org/10.1021/acsnano.3c02511.
(21) May, A. F.; Yan, J.; McGuire, M. A. A Practical Guide for Crystal Growth of van Der Waals Layered Materials. *J. Appl. Phys.* **2020**, *128* (5), 051101. https://doi.org/10.1063/5.0015971.
(22) Pack, J.; Guo, Y.; Liu, Z.; Jessen, B. S.; Holtzman, L.; Liu, S.; Cothrine, M.; Watanabe, K.; Taniguchi, T.; Mandrus, D. G.; Barmak, K.; Hone, J.; Dean, C. R. Charge-Transfer Contacts for the Measurement of Correlated States in High-Mobility $WSe_2$. *Nat. Nanotechnol.* **2024**, *19* (7), 948–954. https://doi.org/10.1038/s41565-024-01702-5.
(23) Yan, J.-Q.; Sales, B. C.; Susner, M. A.; McGuire, M. A. Flux Growth in a Horizontal Configuration: An Analog to Vapor Transport Growth. *Phys. Rev. Mater.* **2017**, *1* (2), 023402. https://doi.org/10.1103/PhysRevMaterials.1.023402.
(24) Park, H.; Li, W.; Hu, C.; Beach, C.; Gonçalves, M.; Mendez-Valderrama, J. F.; Herzog-Arbeitman, J.; Taniguchi, T.; Watanabe, K.; Cobden, D.; Fu, L.; Bernevig, B. A.; Regnault, N.; Chu, J.-H.; Xiao, D.; Xu, X. Observation of Dissipationless Fractional Chern Insulator. *Nat. Phys.* **2026**, 1–7. https://doi.org/10.1038/s41567-025-03167-2.
(25) Soluyanov, A. A.; Gresch, D.; Wang, Z.; Wu, Q.; Troyer, M.; Dai, X.; Bernevig, B. A. Type-II Weyl Semimetals. *Nature* **2015**, *527* (7579), 495–498. https://doi.org/10.1038/nature15768.
(26) Li, P.; Wen, Y.; He, X.; Zhang, Q.; Xia, C.; Yu, Z.-M.; Yang, S. A.; Zhu, Z.; Alshareef, H. N.; Zhang, X.-X. Evidence for Topological Type-II Weyl Semimetal $WTe_2$. *Nat. Commun.* **2017**, *8* (1), 2150. https://doi.org/10.1038/s41467-017-02237-1.
(27) Lin, C.-L.; Arafune, R.; Liu, R.-Y.; Yoshimura, M.; Feng, B.; Kawahara, K.; Ni, Z.; Minamitani, E.; Watanabe, S.; Shi, Y.; Kawai, M.; Chiang, T.-C.; Matsuda, I.; Takagi, N. Visualizing Type-II Weyl Points in Tungsten Ditelluride by Quasiparticle Interference. *ACS Nano* **2017**, *11* (11), 11459–11465. https://doi.org/10.1021/acsnano.7b06179.

(28) Sie, E. J.; Nyby, C. M.; Pemmaraju, C. D.; Park, S. J.; Shen, X.; Yang, J.; Hoffmann, M. C.; Ofori-Okai, B. K.; Li, R.; Reid, A. H.; Weathersby, S.; Mannebach, E.; Finney, N.; Rhodes, D.; Chenet, D.; Antony, A.; Balicas, L.; Hone, J.; Devereaux, T. P.; Heinz, T. F.; Wang, X.; Lindenberg, A. M. An Ultrafast Symmetry Switch in a Weyl Semimetal. *Nature* **2019**, *565* (7737), 61–66. https://doi.org/10.1038/s41586-018-0809-4.

(29) Leahy, I. A.; Lin, Y.-P.; Siegfried, P. E.; Treglia, A. C.; Song, J. C. W.; Nandkishore, R. M.; Lee, M. Nonsaturating Large Magnetoresistance in Semimetals. *Proc. Natl. Acad. Sci.* **2018**, *115* (42), 10570–10575. https://doi.org/10.1073/pnas.1808747115.

(30) Niu, R.; Zhu, W. K. Materials and Possible Mechanisms of Extremely Large Magnetoresistance: A Review. *J. Phys. Condens. Matter* **2021**, *34* (11), 113001. https://doi.org/10.1088/1361-648X/ac3b24.

(31) Liang, T.; Gibson, Q.; Ali, M. N.; Liu, M.; Cava, R. J.; Ong, N. P. Ultrahigh Mobility and Giant Magnetoresistance in the Dirac Semimetal Cd3As2. *Nat. Mater.* **2015**, *14* (3), 280–284. https://doi.org/10.1038/nmat4143.

(32) Xu, S.-Y.; Ma, Q.; Shen, H.; Fatemi, V.; Wu, S.; Chang, T.-R.; Chang, G.; Valdivia, A. M. M.; Chan, C.-K.; Gibson, Q. D.; Zhou, J.; Liu, Z.; Watanabe, K.; Taniguchi, T.; Lin, H.; Cava, R. J.; Fu, L.; Gedik, N.; Jarillo-Herrero, P. Electrically Switchable Berry Curvature Dipole in the Monolayer Topological Insulator $WTe_2$. *Nat. Phys.* **2018**, *14* (9), 900–906. https://doi.org/10.1038/s41567-018-0189-6.

(33) Song, T.; Jia, Y.; Yu, G.; Tang, Y.; Uzan, A. J.; Zheng, Z. J.; Guan, H.; Onyszczak, M.; Singha, R.; Gui, X.; Watanabe, K.; Taniguchi, T.; Cava, R. J.; Schoop, L. M.; Ong, N. P.; Wu, S. Unconventional Superconducting Phase Diagram of Monolayer $WTe_2$. *Phys. Rev. Res.* **2025**, *7* (1), 013224. https://doi.org/10.1103/PhysRevResearch.7.013224.

(34) Ginzburg, V. L. The Problem of High-Temperature Superconductivity. II. *Sov. Phys. Uspekhi* **1970**, *13* (3), 335. https://doi.org/10.1070/PU1970v013n03ABEH004256.

(35) Crépel, V.; Fu, L. Spin-Triplet Superconductivity from Excitonic Effect in Doped Insulators. *Proc. Natl. Acad. Sci.* **2022**, *119* (13), e2117735119. https://doi.org/10.1073/pnas.2117735119.

(36) Massalski, T. B.; Okamoto, H.; International, A. S. M. *Binary Alloy Phase Diagrams*; Binary Alloy Phase Diagrams; ASM International, 1990.

(37) Zhu, Z.; Lin, X.; Liu, J.; Fauqué, B.; Tao, Q.; Yang, C.; Shi, Y.; Behnia, K. Quantum Oscillations, Thermoelectric Coefficients, and the Fermi Surface of Semimetallic $WTe_2$. *Phys. Rev. Lett.* **2015**, *114* (17), 176601. https://doi.org/10.1103/PhysRevLett.114.176601.

(38) Kwon, H.; Jeong, T.; Appalakondaiah, S.; Oh, Y.; Jeon, I.; Min, H.; Park, S.; Song, Y. J.; Hwang, E.; Hwang, S. Quasiparticle Interference and Impurity Resonances on $WTe_2$. *Nano Res.* **2020**, *13* (9), 2534–2540. https://doi.org/10.1007/s12274-020-2892-8.

(39) Song, H.-Y.; Lü, J.-T. Single-Site Point Defects in Semimetal $WTe_2$: A Density Functional Theory Study. *AIP Adv.* **2018**, *8* (12), 125323. https://doi.org/10.1063/1.5057723.

(40) Chen, W.-H.; Kawakami, N.; Lin, J.-J.; Huang, H.-I.; Arafune, R.; Takagi, N.; Lin, C.-L. Noncentrosymmetric Characteristics of Defects on $WTe_2$. *Phys. Rev. B* **2022**, *106* (7), 075428. https://doi.org/10.1103/PhysRevB.106.075428.

**Supplementary Information**

**Record magnetoresistance, superconductivity, and fermiology in $WTe_2$**

Delgado et al

# Contents

## SI1. Scanning tunneling microscopy images and defect species analysis

The defect density reported in the main text is calculated from STM topography images taken from multiple macroscopic regions across two cleaves of an HFT-grown $WTe_2$ crystal. This batch of crystals was not included in Fig. 1e. Fig. S1.1a shows an image from the same cleave but from a different region to that in Fig. 1f, and Fig. S1.1b shows an image from a separate cleave, demonstrating varying densities within the same crystal. In total, 172 defects were recorded over an area of 115,185 nm$^2$, yielding a defect density of $(1.5 \pm 0.1) \times 10^{11}\ \mathrm{cm}^{-2}$. The uncertainty assumes independent Bernoulli trials associated with each measured atom, though we note there is likely larger variations across crystals, as indicated by the spread in crystal quality shown in Fig. 1e. The density was converted to defect per unit cell in the main text with lattice constants measured by the STM, thereby negating any potential error from piezo miscalibration.

Two primary kinds of defects were observed in this crystal, denoted "A" (white circles) and "B" (blue triangles) in Fig. 1.1b. Zoomed in images of defect "A" are shown in Figs. 1.1c-d, showing opposite contrast for different signs of $V_b$. Its appearance at $V_b > 0$ closely matches that of Te vacancies previously identified in the literature[38]. Under certain tip conditions, a subset of these defects exhibit an extra central bright spot (white arrows in Fig. S1.1a). We postulate that those with a bright spot are vacancies on the uppermost layer of Te atoms, while those without are on the layer of Te atoms below W. Additionally, faint patches with a similar shape can be seen in some images (gray circles in Fig. S1.1a). We believe these to be Te vacancies on a buried $WTe_2$ layer and thus have excluded them from the per-layer defect count. Figs. S1.1e-f show zoomed in images of defect "B", which has a similar contrast switching behavior with $V_b$. Its appearance resembles that of a Te substitution.[39,40] Overall, across all images in which defects"A" and "B" can be clearly distinguished, "A" appeared $13 \pm 6$ times as frequently as "B".

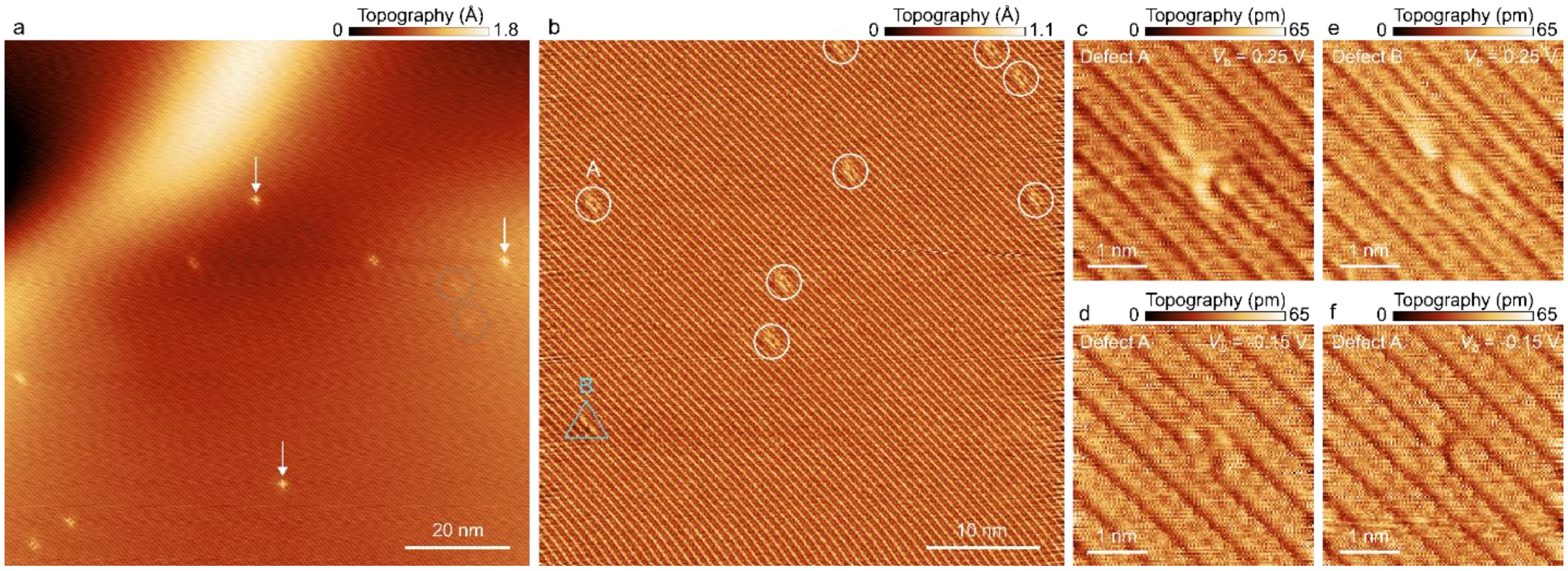

**Fig. S1.1.** Additional STM images of the same $WTe_2$ crystal as Fig. 1f. (a) 100x100 nm$^2$ STM image of a different macroscopic region of the crystal shown in Fig. 1f. White arrows point to defects with an extra central bright spot. Gray circles denote defects on the underlying layer. (b) 46x46 nm$^2$ STM image of this crystal from a separate cleave. White circles and blue triangle denote the two dominant defect species found on this crystal, labeled "A" and "B". (c-d) Zoomed in STM images of defect "A" at $V_b$ = 0.25 V (c) and $V_b$ = -0.15 V (d), showing opposite contrasts with opposite $V_b$ and an appearance consistent with Te vacancies. (e-f) same as c-d but for defect "B", showing similar contrast switching behavior and an appearance resembling Te substitutions.

To make a connection to the crystals analyzed in Fig. 1e, we show in Fig. S1.2a a representative STM topograph of a cleaved crystal belonging the batch plotted as blue squares. Here, across 82,500 nm$^2$, the density of defect "A" was $(3.6 \pm 0.7) \cdot 10^{10}$ cm$^{-2}$, whereas no defects of type "B" were found. Instead, we see a third defect species "C" (green diamonds), though we are unable to obtain its density due to vibrational noise. We observed similar defects in a crystal from a third batch, which was also not plotted in Fig. S1e. A topograph of it is shown in Fig. S1.2b. Here, from a total area of 139,025 nm$^2$, we recorded densities of $(5 \pm 2) \cdot 10^9$ cm$^{-2}$ for defect "A" and $(2.5 \pm 0.4) \cdot 10^{10}$ cm$^{-2}$ for defect "C"; defect "B" is again not present. Figs. S1.2c-d show zoomed in images of defect "C". The bias dependence confirms that it is distinct from defects "A" and "B". Because these defects do not interrupt the observed atomic chains, in contrast to defects "A" and "B", we believe these to be defects on W sites. Since both crystals were annealed post-growth to aid flux separation, we hypothesize that defect "C" is thermally induced. On the other hand, the differences in densities of defects "A" and "B" can possibly be attributed to the inherent variation between individual crystals and between batches.

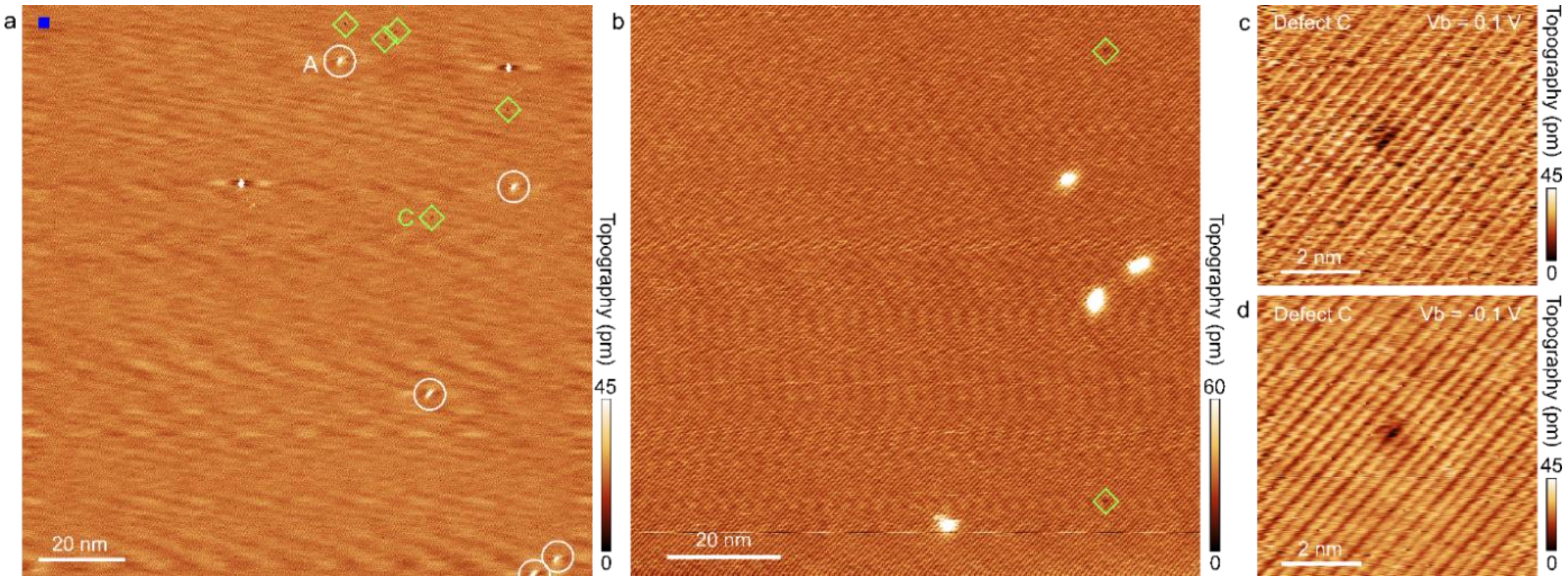

**Fig. S1.2.** STM images of HFT-grown $WTe_2$ crystals from other batches. (a) 132x132 nm$^2$ STM image of a crystal from the batch plotted as blue squares in Fig. 1e. White circles denote defect "A". A new defect species observed here is marked by green diamonds and labeled "C". (b) 100x100 nm$^2$ STM image of a crystal from a separate batch, in which defect "C" are also present. (c-d) Zoomed in images of a defect "C" taken at $V_b$ = 0.1 V (c) and $V_b$ = -0.1 V (d). These are likely defects on W sites as they do not interrupt the atomic chains.

**Table S2.1. Tunneling conditions for STM images.**

| Fig. | 1f | S2.1a | S2.1b | S2.1c | S2.1d | S2.1e | S2.1f |
|---|---|---|---|---|---|---|---|
| $V_b$ (V) | 0.3 | 0.3 | 0.25 | 0.25 | -0.15 | 0.25 | -0.15 |
| $I$ (pA) | 20 | 20 | 100 | 100 | 100 | 100 | 100 |

| S2.2a | S2.2b | S2.2c | S2.2d |
|---|---|---|---|
| 0.25 | 0.3 | 0.1 | -0.1 |
| 250 | 20 | 20 | 20 |

## SI2. Monolayer device fabrication

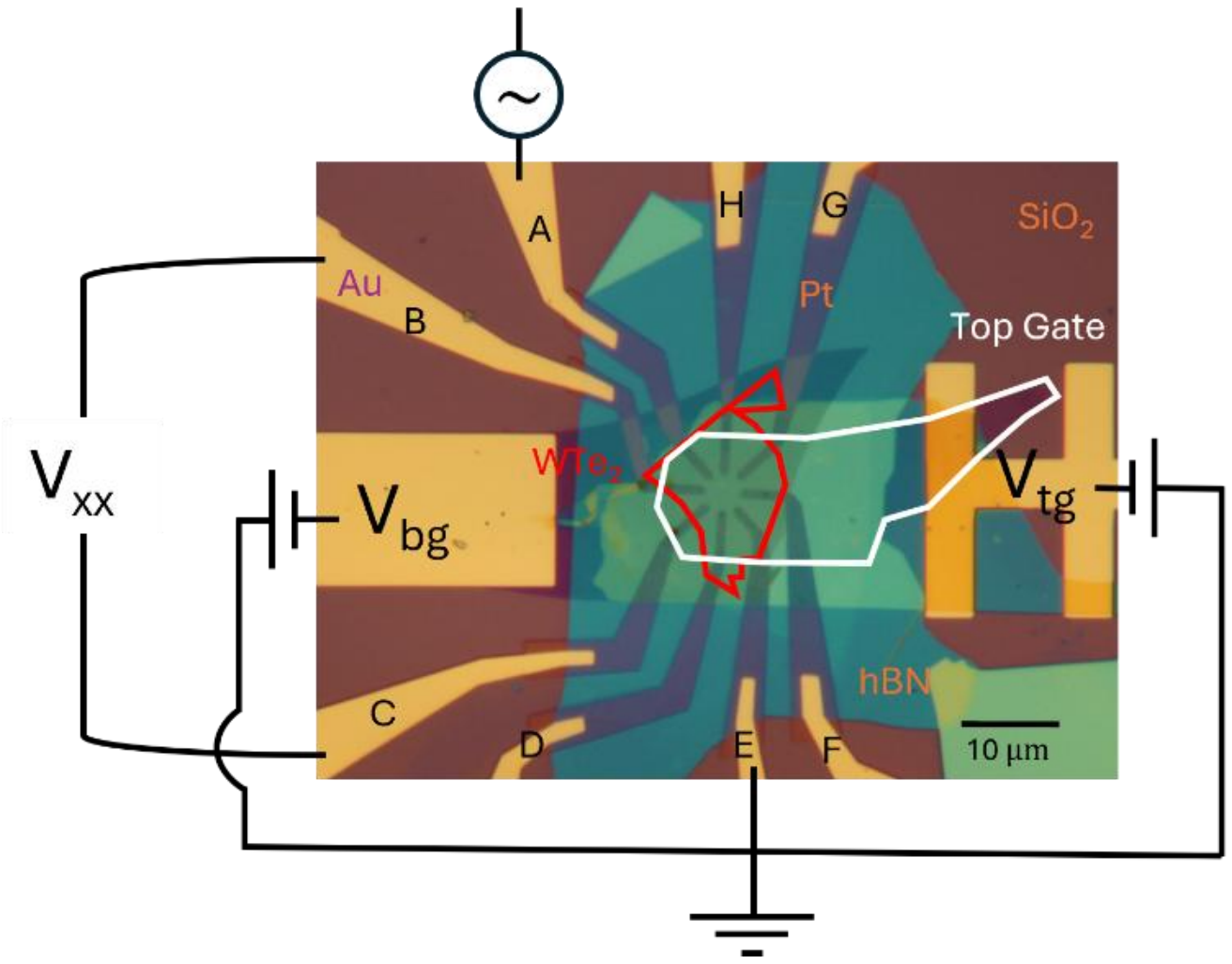

**Fig. S2.1.** Optical image of device 1 with wiring diagram. The red outline denotes the boundary of the monolayer $WTe_2$ flake. Electrodes not shown connected are left floating.

Figure S2.1 shows an optical image of device 1 with a wiring diagram showing the four-terminal configuration used for the resistivity measurements plotted in figures 2 and 3. To find $n_e$ and $D$ as used in the main text we apply the following transformations to the applied bottom ($V_{bg}$) and top ($V_{tg}$) gate voltages.

$$n_e = \left(C_{bg}V_{bg} + C_{tg}V_{tg}\right)/e$$

$$D_\perp = C_{bg}V_{bg} - C_{tg}V_{tg} \text{ ,}$$

where $C_{bg}$ and $C_{tg}$ are the capacitance per unit area of the bottom and top gates respectively and $e$ is the charge of the electron. The capacitances are calculated by linearly fitting the dispersion of the SdH oscillations with gate voltage and magnetic field yielding $C_{tg}$ = 6.4 fFμm$^{-2}$ and $C_{tg}$ = 1.3 fFμm$^{-2}$.

## SI3. Quantum oscillations and mean free path in the monolayer electron-doped metallic state

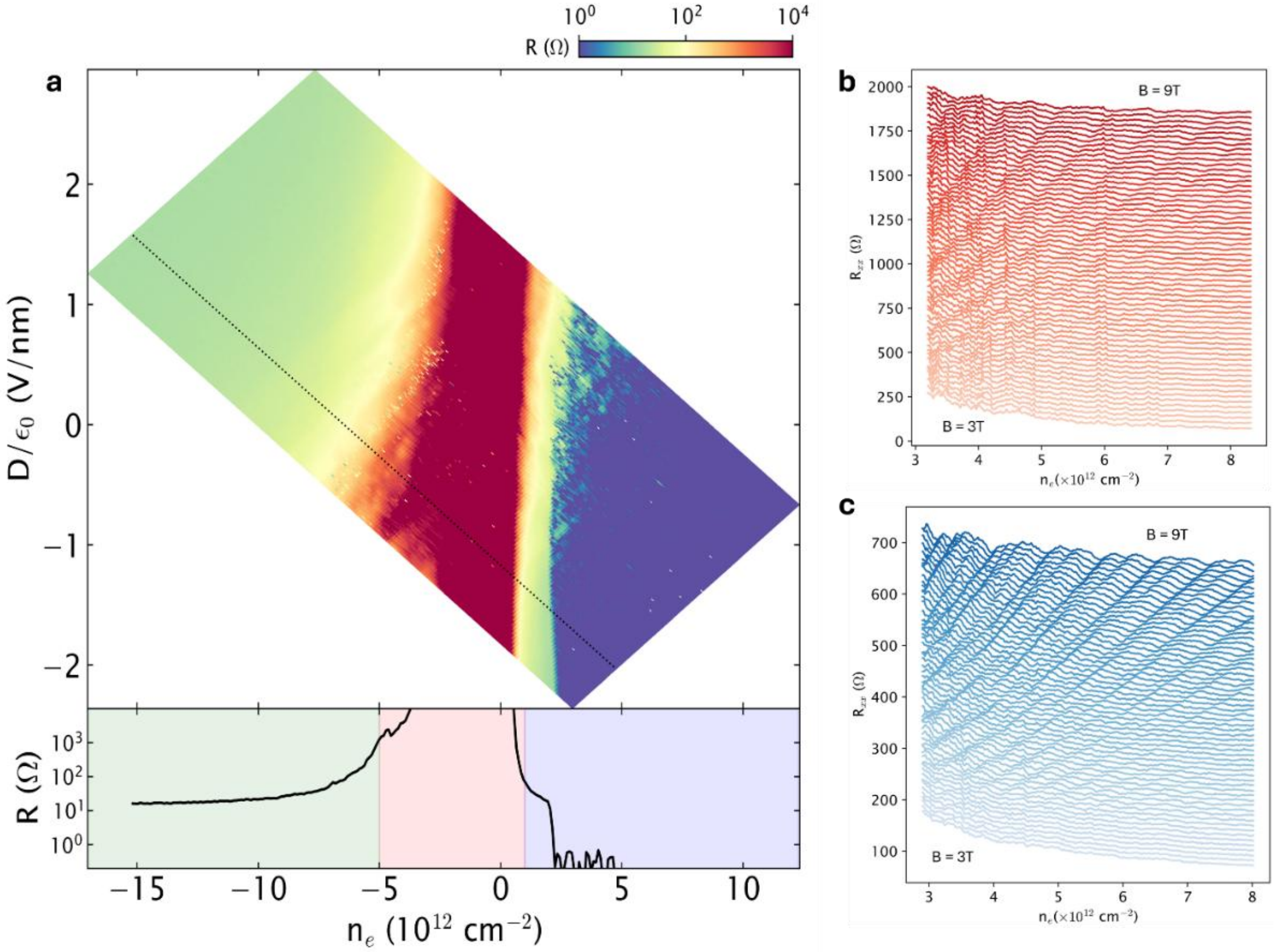

**Fig. S3.1.** a) Gate-gate map of the four-terminal resistance at $B = 0$ and base temperature, as shown in Fig.4. Below is a line cut along the dotted black line (tracking a single gate), with green shading indicating the hole metal, red the insulating state and blue the superconducting state. b) Waterfall plot of the transverse (not symmetrized) resistance as a function of doping for a series of magnetic fields at $D = 0$ V/nm. Each line is sequentially offset by 15 Ω. The SdH oscillations are manifest, but their amplitude is no larger than ~20%. c) As for (b) but at $D = 1.25$ V/nm and with sequential offset 5 Ω.

The absolute value of the sheet resistivity, $\rho$, is uncertain up to a numerical factor because the $WTe_2$ is not shaped in a simple geometry. Nevertheless, taking that factor to be unity, we estimate the normal state resistivity to be $\rho \approx 60\Omega$ at $n_e = 10^{13}$ cm$^{-2}$ from Fig. 3b. From this we can make a rough estimate of the transport mobility, $\mu \approx (\rho n e)^{-1} = 1.0 \times 10^4$ cm$^2$V$^{-1}$s$^{-1}$. We can also use the onset of SdH oscillations at $B_{SdH} \approx 3.4$ T (at which the period of the cyclotron orbit roughly equals the small-angle scattering time at the Fermi level) to estimate a quantum mobility (corresponding to small-angle scattering) of $\mu_q \approx \frac{1}{B_{SdH}} \approx 3 \times 10^3$ cm$^2$V$^{-1}$s$^{-1}$. This is smaller than $\mu$ as expected (but more closely matching than in the bulk $WTe_2$.) Assuming parabolic dispersion and degeneracy of 4, using $\mu \approx 1.0 \times 10^4$ cm$^2$V$^{-1}$s$^{-1}$ gives a mean free path estimate of

$$\lambda = 2\hbar(\pi^3 n_e)^{1/2}\mu/e \approx 2\ \mu\text{m}.$$

This is larger than the superconducting coherence length of ~90 nm (see Fig. 3d and Fig. S5.1b) and thus consistent with the superconductor at high $n_e$ being in the clean limit.

## SI4. Valley splitting

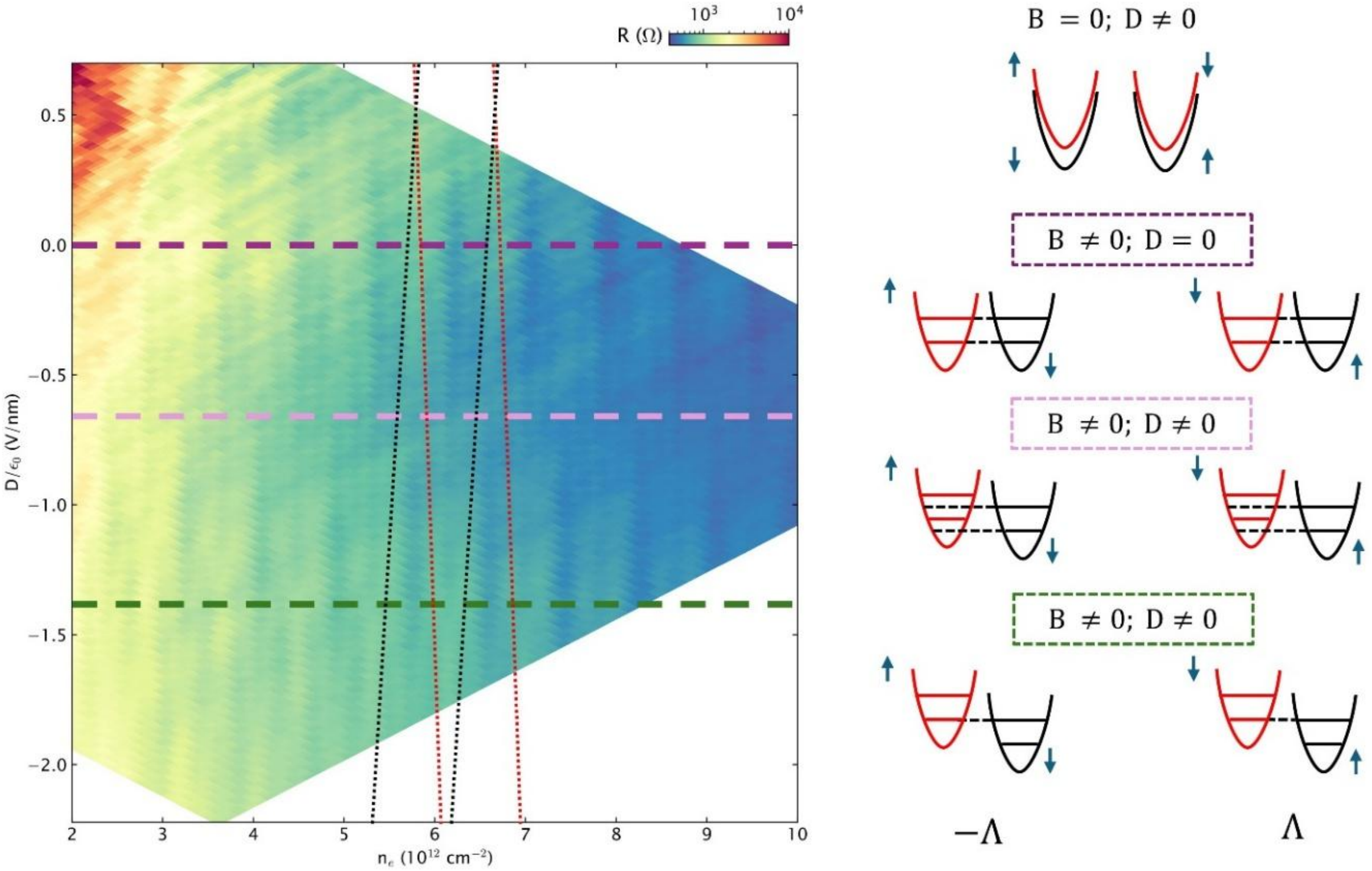

**Fig. S4.1.** (left) Zoom in on the SdH oscillations at 9 T on the electron-doped side in Fig 2a. The red and black dotted lines, with equal and opposite slope, indicate the linear motion with displacement field of resistance peaks corresponding to Landau levels associated with the two valleys. The situation at each of the values of $D_{\perp}$ indicated by the horizontal dashed lines is shown in the schematics on the right.

Fig. S4.1 provides an enlarged view of the dual-gate map from Fig. 2a. Near $D = 0$ V/nm, the Shubnikov–de Haas (SdH) oscillations exhibit the expected period for fourfold degenerate carriers at the fermi level. As the displacement field ($D_{\perp}$) increases, the oscillation amplitude diminishes, nearly vanishing at ~ -0.75 V/nm before reappearing with a π phase shift at ~ -1.4 V/nm. Closer inspection reveals that the single resistance peaks at $D_{\perp} = 0$ V/nm split into two distinct branches that diverge linearly with $D_{\perp}$. This evolution is consistent with a combination of Zeeman spin-splitting (set by the applied external magnetic field) and electric-field-induced valley splitting of the spins. Based on the slope of these resistance peaks with $D_{\perp}$ and assuming a density of states in the conduction band of $3.7 \times 10^{11}\ \text{cm}^{-2}\text{meV}^{-1}$, we calculate an inter-valley band shift of 1.4 meV per V/nm. This is much smaller than the 72 meV at 1 V/nm calculated in Ref. 32.

## SI5. BCS theory comparison

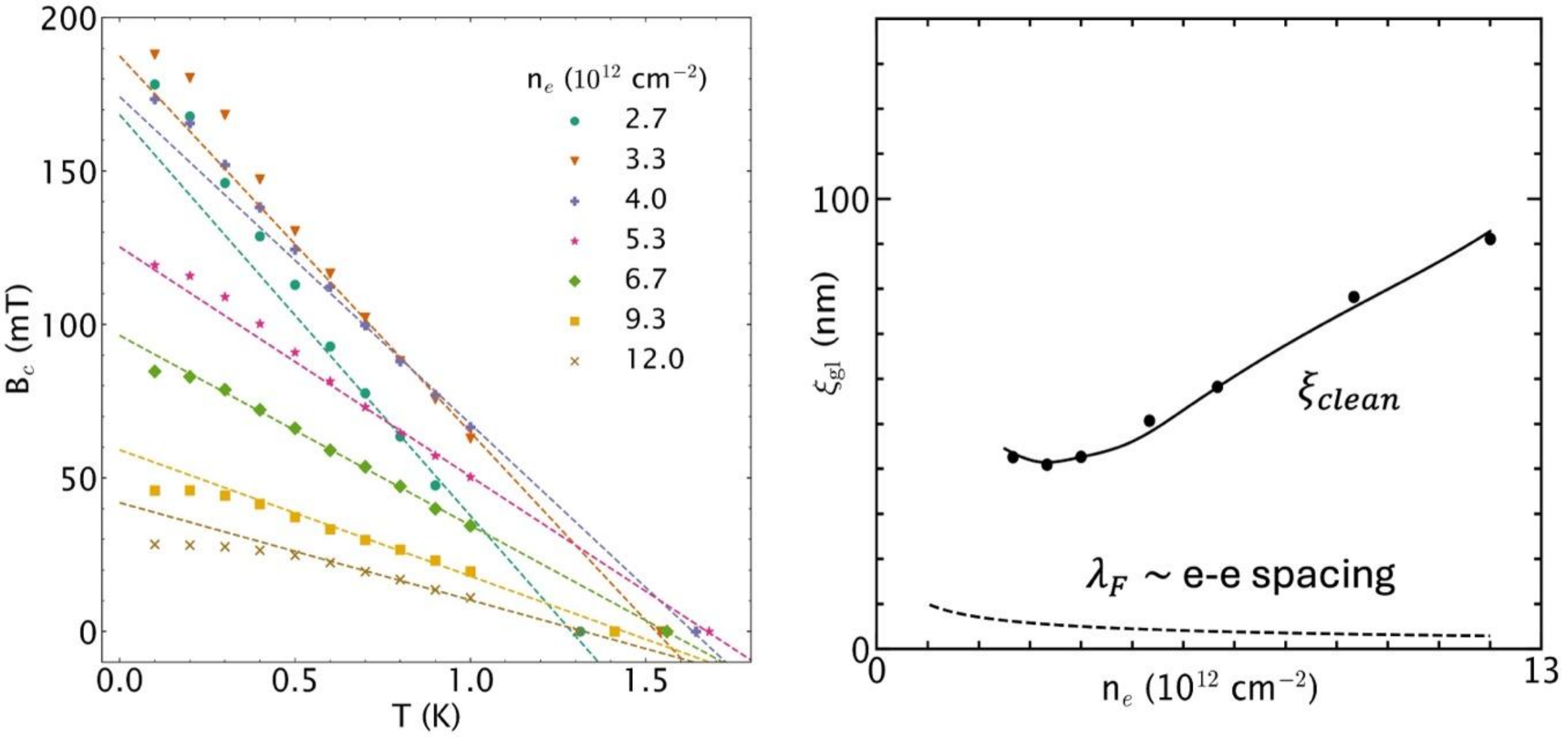

**Fig S5.1. Coherence Length and Comparison to Clean Limit BCS.** a) Critical perpendicular magnetic field ($B_c$) versus Temperature (T) at various electron dopings. The dashed lines are linear fits to the discrete points for the points above 0.5 K. b) Plot of the Ginzburg-Landau (GL) coherence length extracted from the slopes of the fits in (a) (black dots) with a fit from clean limit expectation from BCS theory (solid black line) and the interelectron spacing (black dashed line).

In figure (S.5.1a) we plot the temperature dependence of $B_c$ at various electrons dopings. At temperatures above ~0.5 K it is close to linear and can be fit to the GL relation, $B_c(T) = \frac{\phi_0}{2\pi\xi_{GL}^2}\left(1 - \frac{T}{T_c}\right)$, where $T_c$ is the zero-field critical temperature, $\phi_0$ is the 2e magnetic flux quantum, and $\xi_0$ is the GL coherence length. Extracting $\xi_{GL}$ from the fitted slopes (corresponding to $\xi_{GL} = \sqrt{\phi_0/2\pi B_c(T\to 0)}$, we get the dependence on $n_e$ plotted in the inset in Fig. 3d.

In a clean superconductor in BCS theory the coherence length is proportional to the ratio of the Fermi velocity of the electrons in the parent state and the gap size. Assuming the conduction bands of monolayer $WTe_2$ have a parabolic dispersion, this leads to the following relation:

$$\xi_{BCS} = \frac{\hbar v_F}{\pi\Delta_0} = \frac{2}{\pi^{3/2}(1.76\, k_b)D(E)}\frac{\sqrt{n_e}}{T_c}$$

where $k_b$ is the Boltzmann constant, $D(E)$ is the density of states in the conduction band, and $n_e$ is the density of electron carriers. The only free parameter is $D(E)$ and finding the best fit to the coherence length calculated from the slopes fitted in S5.1a gives a density of states of $D_0(E) = 1.2 \times 10^{12}\mathrm{cm}^{-2}\mathrm{meV}^{-1}$ which compares to the density of states measured from compressibility measurements of $D_0(E) = 3.7 \times 10^{11}\mathrm{cm}^{-2}\mathrm{meV}^{-1}$ [18]. The good fit to the doping dependence of the coherence length attained from the $B_c$ vs $T_c$ dependence and the order of magnitude match to the experimentally measured density of states is further evidence that the superconductor state observed in device 1 is in the clean limit.

Furthermore, plotting the average interelectron distance makes it clear that the coherence length far exceeds the interelectron spacing and therefore the fermi wavelength of the charge carriers meaning the weak coupling assumption in the BCS framework employed above is self-consistent.

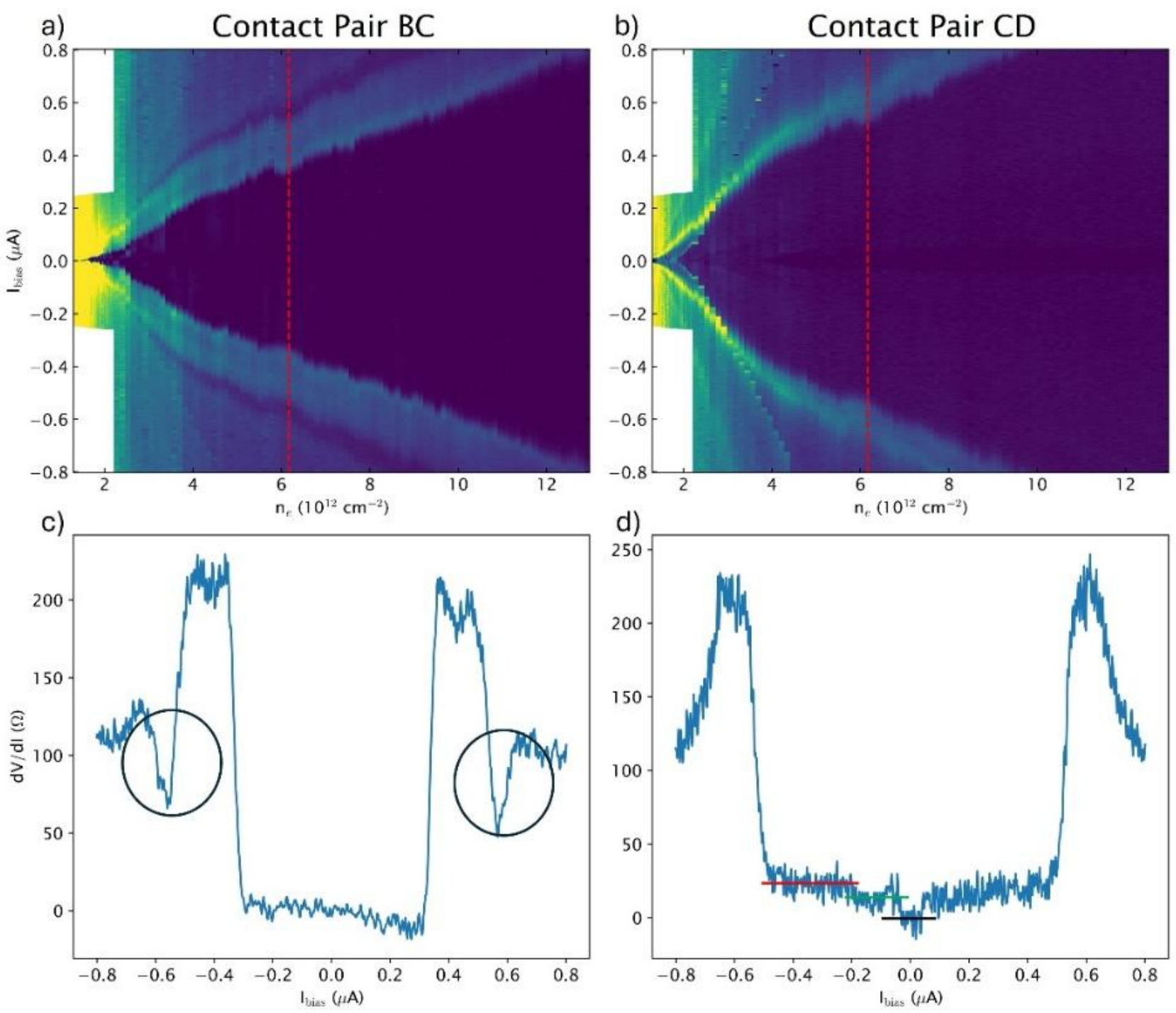

**Fig S6.1. Contact pair variations in critical current of the SC.** Color map of differential resistance ($dV/dI$) as a function of DC current bias ($I_{bias}$) and carrier concentration ($n_e$) for contact pairs BC (a,c) and CD (b,d) as labelled in Fig S3.1, with line cuts taken at $n_e = 6.1 \times 10^{12} \text{cm}^{-2}$ along the dashed red lines in (a) and (b).

## SI6. Critical current measurements

Although the general trend of the critical current with doping is similar between different contacts in device 1, there are significant variations in the details. For example, Fig. S6.1 compares two contact pairs (BC and CD). For pair BC the critical current $I_c$ is smaller and the differential resistance shows a dip above $I_c$ that is not observed for pair CD.

## SI7. Analysis of bulk crystal transport

Four different batches of $WTe_2$ crystals were examined by checking the RRR and mobility via electric transport measurements. Figure S1 summarizes the all resistance versus temperature and field (B//c) data for 32 samples. The extracted RRR and mobility are used to construct Figure 1d. Crystal orientation relative to the field is verified by quantum oscillation frequency. RRR is calculated using $R(300K)/R_{base}$, where the base temperature is 1.7 K or 100 mK. For a system like $WTe_2$ with nearly compensated charge carrier , $n_h \approx n_e$ , $\rho_{xx} = \frac{1}{e}\frac{n_e\mu_e + n_h\mu_h + (n_e\mu_h + n_h\mu_e)\mu_h\mu_e B^2}{(n_e\mu_e + n_h\mu_h)^2} = \rho_{xx,0}(1 + \mu_h\mu_e B^2) = \rho_{xx,0}\left(1 + \overline{\mu}^2 B^2\right)$. The average transport mobility for both hole and electron, shown in each R(H) plot in Figure S7.1, is simply calculated by the magnetoresistance at a high field $\overline{\mu} = \sqrt{MR}/H$, where MR= $\frac{R(H)}{R(0)} - 1$.

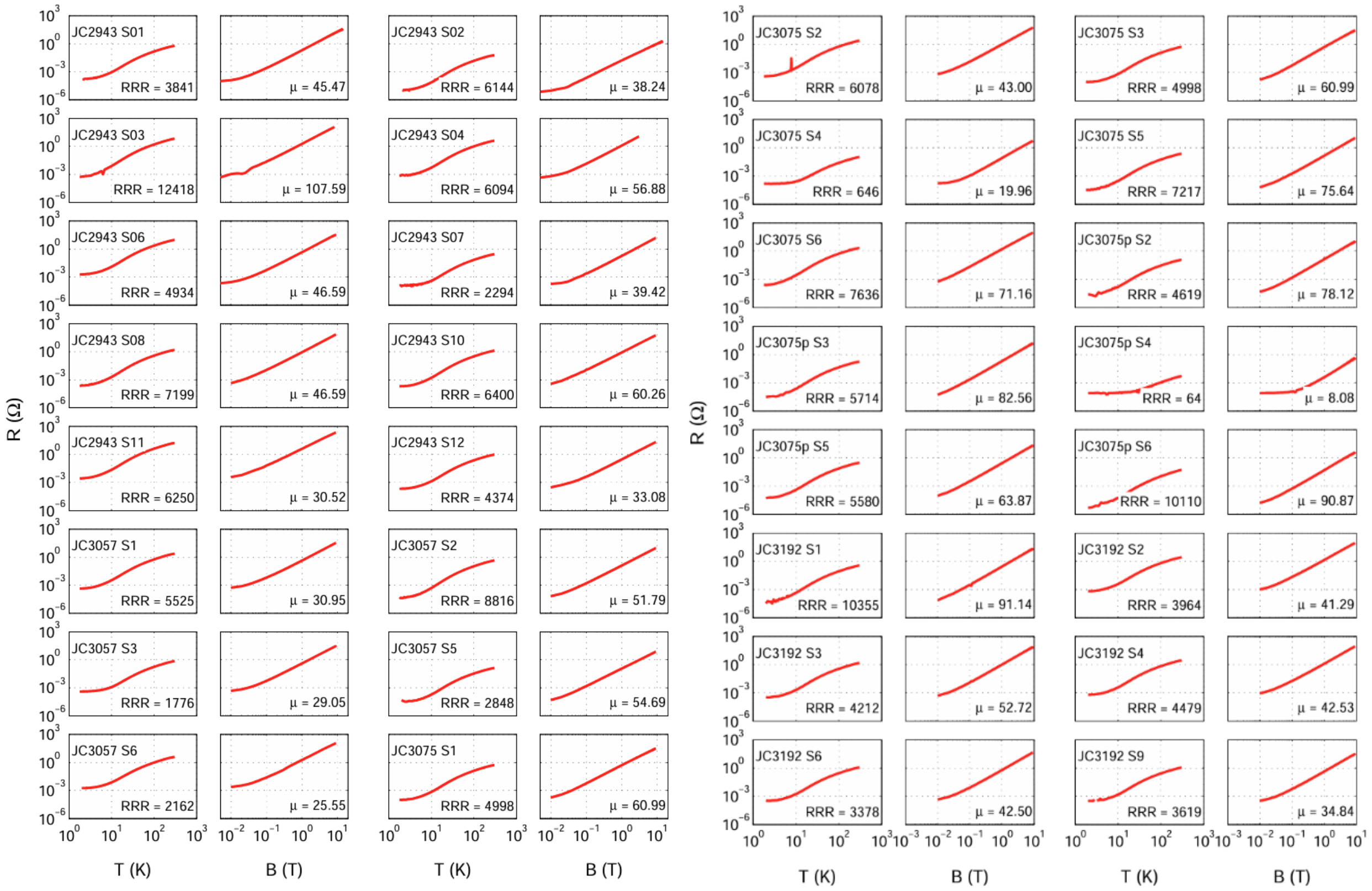

**Figure S7.1**. Resistance versus temperature and field plots for 32 samples. RRR in the text is unitless while the unit for μ in the figure is $m^2V^{-1}s^{-1}$. The crystal orientation relative to the field (B//c) is verified by quantum oscillation frequencies.

The value of exact hole and electron density and mobility can also be directly obtained from longitudinal and transverse resistivity, $\rho_{xx}$ and $\rho_{xy}$, using two-band fitting. For $WTe_2$, the longitudinal and transverse resistivity evolves under field as

$$\rho_{xx} = \frac{1}{e}\frac{n_e\mu_e + n_h\mu_h + (n_e\mu_h + n_h\mu_e)\mu_h\mu_e B^2}{(n_e\mu_e + n_h\mu_h)^2 + (n_h - n_e)\mu_h^2\mu_e^2 B^2}$$

$$\rho_{xy} = \frac{B}{e}\frac{-n_e\mu_e^2 + n_h\mu_h^2 + (n_h - n_e)\mu_h^2\mu_e^2 B^2}{(n_e\mu_e + n_h\mu_h)^2 + (n_h - n_e)^2\mu_h^2\mu_e^2 B^2}$$

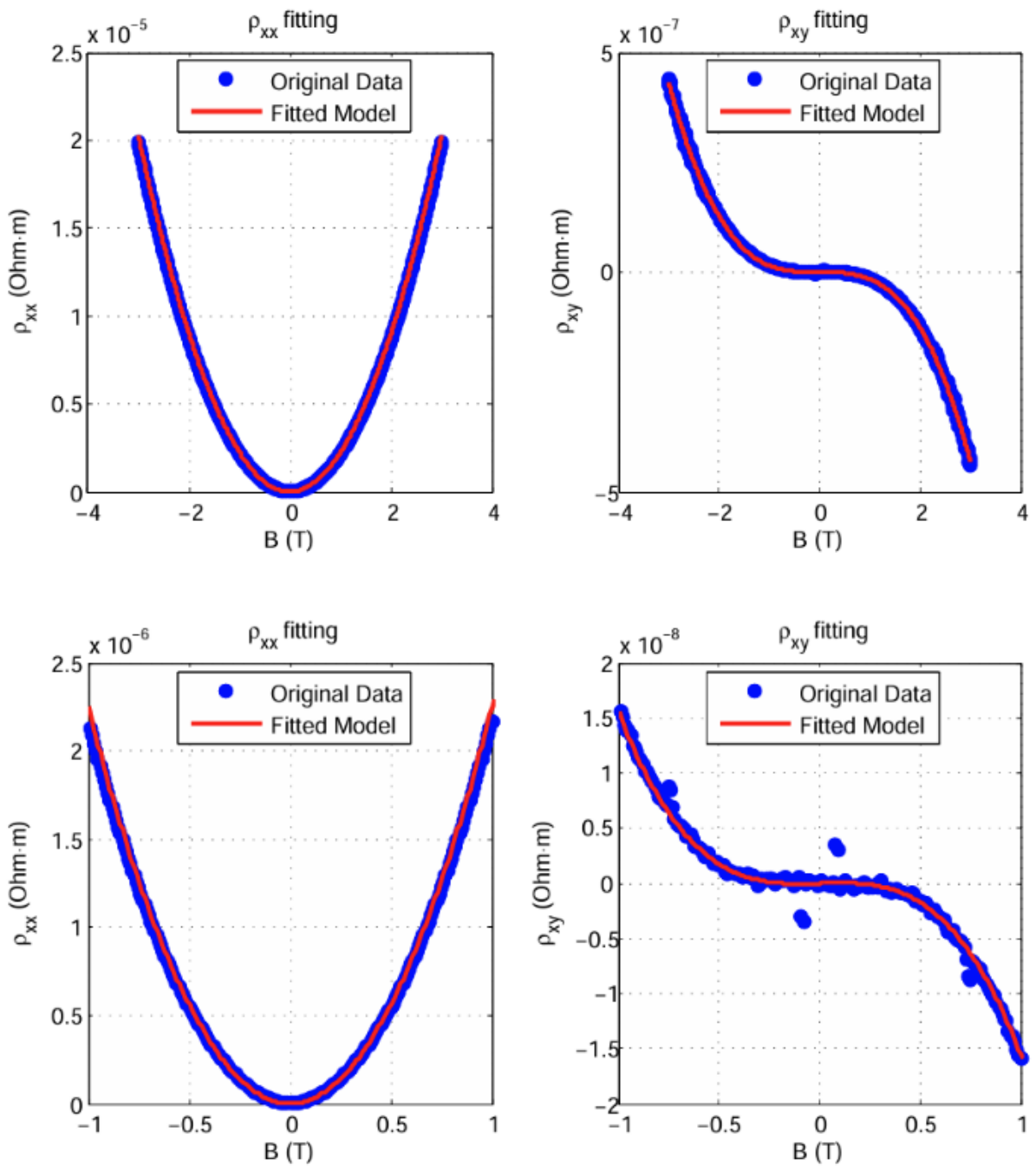

**Figure S7.2.** Two band fitting of $\rho_{xx}$ and $\rho_{xy}$ for JC2943 S4, shown in both high field and low field

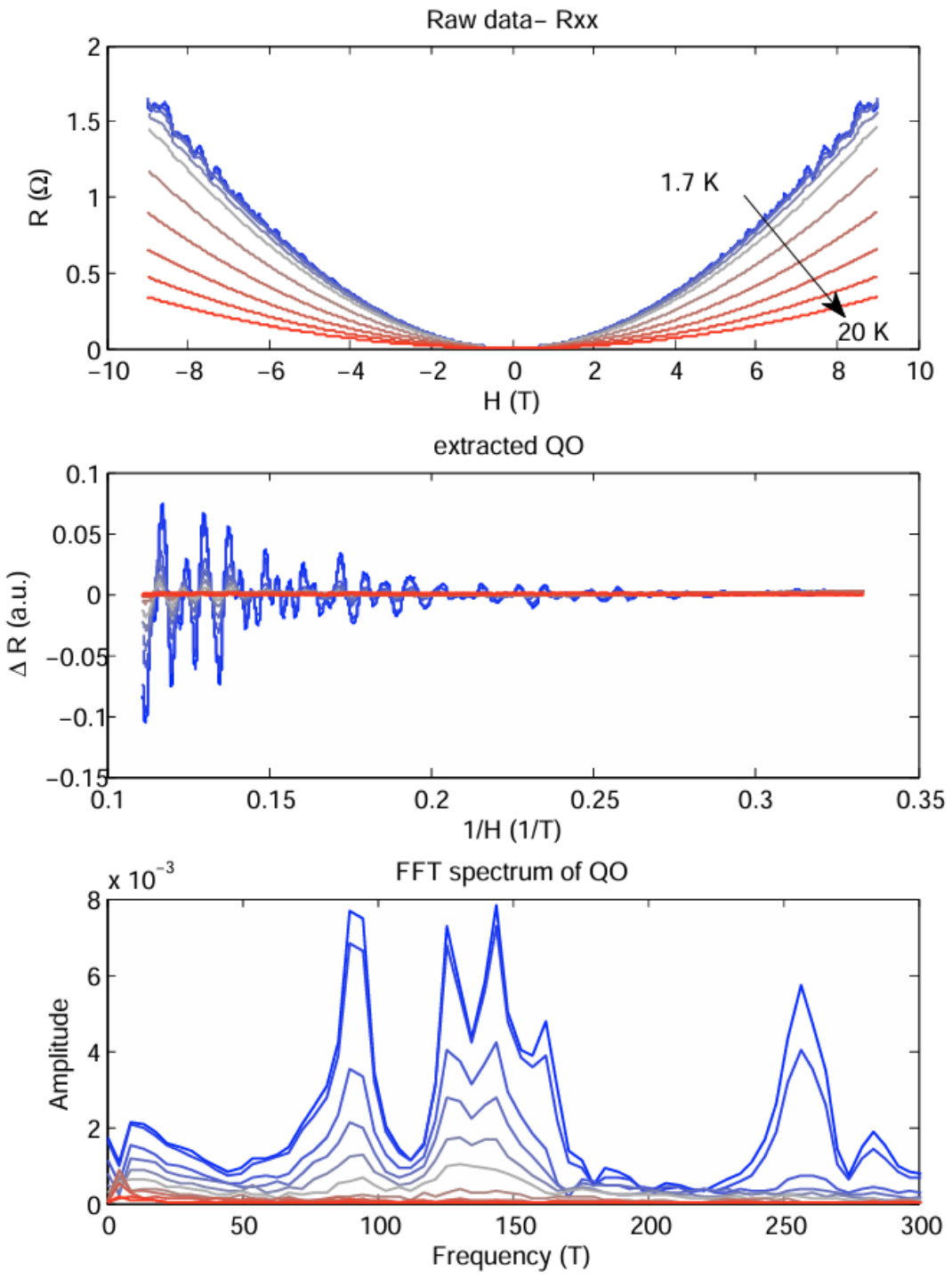

**Figure S7.3.** Magnetoresistance, quantum oscillation and FFT spectrum for JC3092 S7 measured at various temperatures

Figure S7.2 shows an example of simultaneous fitting of $\rho_{xx}$ and $\rho_{xy}$ for with the constraint that the resistivity at B=0 has to be matched exactly. The result shows electron mobility and density to be $\mu_e =$ 570536 cm$^2$ /V s , $n_e$=7.9677 × $10^{19} cm^{-3}$ and hole mobility and density to be $\mu_h$=578322 cm$^2$ /V s, $n_h$=7.9657× $10^{19} cm^{-3}$. The electron density is larger than the hole density only by 0.025%. The fitting also confirms the overall high mobility for both the electron and holes. Their values are similar and close to the estimated average value of 568800 cm$^2$ s calculated from MR directly. Therefore, the MR solely serves as a good probe to characterize $WTe_2$ mobility in the bulk.

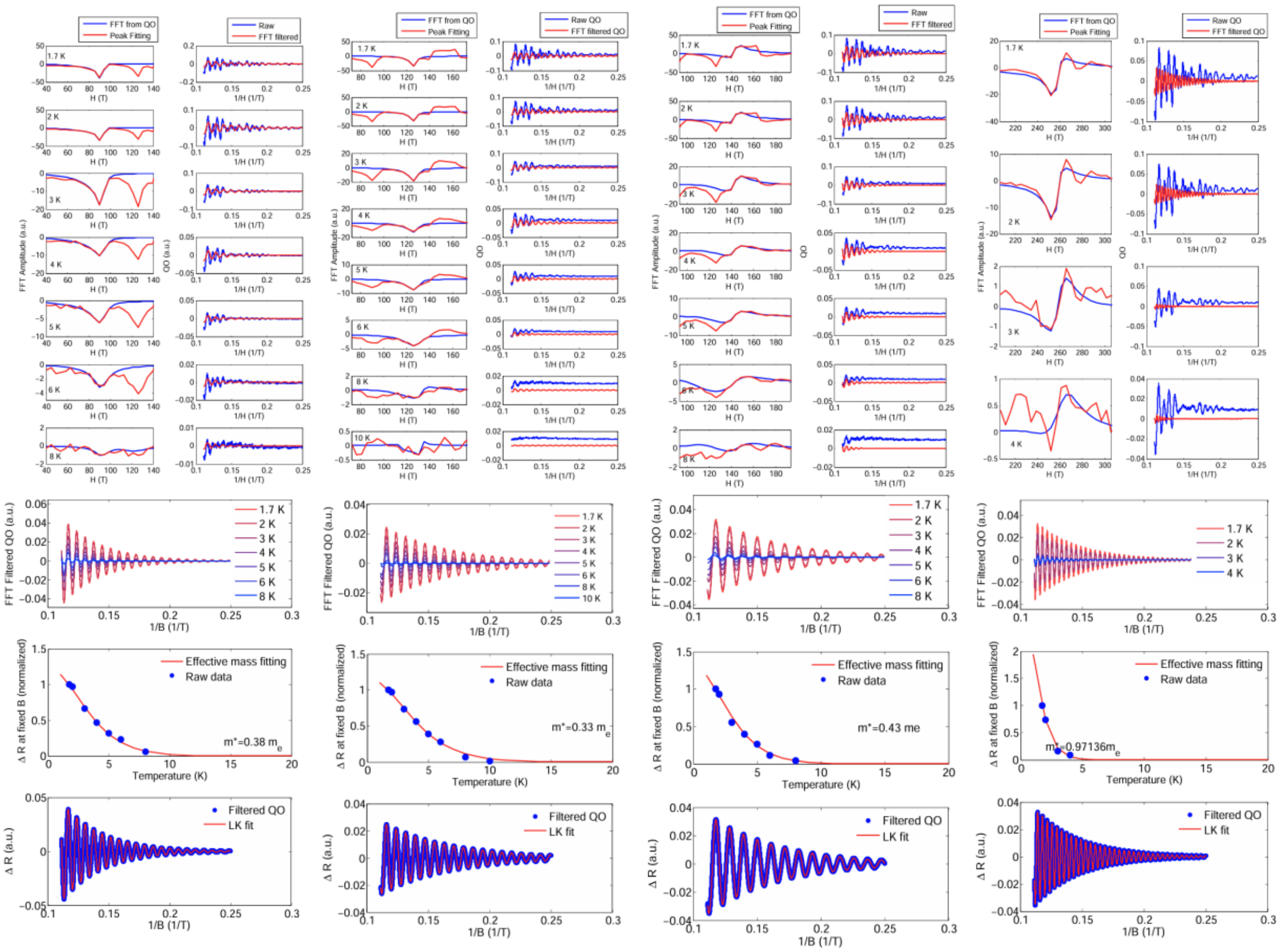

**Figure S7.4.** Quantum mobility analysis for $WTe_2$. For each column, top series of plots, peak fitting FFT spectrum at the designated frequency and the singled-out oscillation at each temperature. Bottom series of plots, top, oscillation at each temperature, mid, fitting oscillation amplitude at fixed field for $R_T$ term, bottom, fitting of the $R_D$ term at the base temperature.

To examine the quantum mobility, quantum oscillations are extracted from the magnetoresistance (Fig S7.3) at various temperature using $\Delta R = R(B) - R_{background}(B)$, where $R_{background}(B) \propto B^2$ is the perfect parabolic background MR. The extracted oscillations are plotted in mid figure in Figure S7.3. From there, fast Fourier transform is performed and shows mainly four peaks at 92 T, 126 T, 143 T, and 256 T in the bottom Figure of Figure S7.3, which is consistent with the previous reports. For each frequency, the corresponding Shubnikov–de Haas (SdH) oscillations is described by the Lifshitz–Kosevich (LK) theory

$$\frac{\Delta R}{R} \propto \frac{1}{\sqrt{B}} R_T R_D \cos\left(2\pi\left(\frac{F}{B}\right) + \phi\right)$$

Here, $R_T = \frac{\frac{\alpha m^* T}{B}}{\sinh\left(\frac{\alpha m^* T}{B}\right)}$ is the thermal damping factor. $R_D = \exp\left(\frac{\alpha m^* T_D}{B}\right)$ is the Dingle damping factor. $\alpha = 2\pi^2 k_B m_e / e\hbar = 14.69\mathrm{T/K}$. $m^*$ is the effective electron mass relative to the free electron mass $m_e$.

However, the presence of four frequencies poses a challenge on the direct fitting of the resistance oscillations. Oscillation needs to be first separated at each frequency. To do so, peak fitting is performed in the FFT spectrum using Lorentzian peak form near the corresponding peak at each temperature. The inverse FFT is subsequently done to reconstruct quantum oscillations arising from the corresponding pocket, and its overall temperature dependence. The result of sample JC3192-S7 is shown in Figure S7.4 to exemplify the process.

From the temperature dependence, we choose a fixed field $B'$ that correponds to the local maximum of the oscillation. Using the amplitude at $B'$, we fitted the temperature dependence with the $R_T = \frac{\frac{\alpha m^* T}{B}}{\sinh\left(\frac{\alpha m^* T}{B}\right)}$, and obtained the effective mass for electron on the corresponding Fermi pocket.

Knowing the effective mass, we finally perform the complete LK fitting on the quantum oscillations at the base temperature, as shown in the bottom figure of each column in Fig S7.4. The fitting yields Dingle temperature $T_D$, from which quantum mobility can be calculated using $\mu_q = \frac{e\hbar}{2\pi k_B m^* T_D}$.

The same process is applied for all four frequencies as shown in Figure S7.4, and then repeated for 8 samples with varying transport mobility. This allows us to thereby construct a summary plot of quantum mobility and effective mass versus transport mobility for all four Fermi pockets. The overall consistent effective mass shows the validity of the fitting. The value of quantum mobility is also consistent with the fact that oscillation onsets around 2-3 T for all samples.

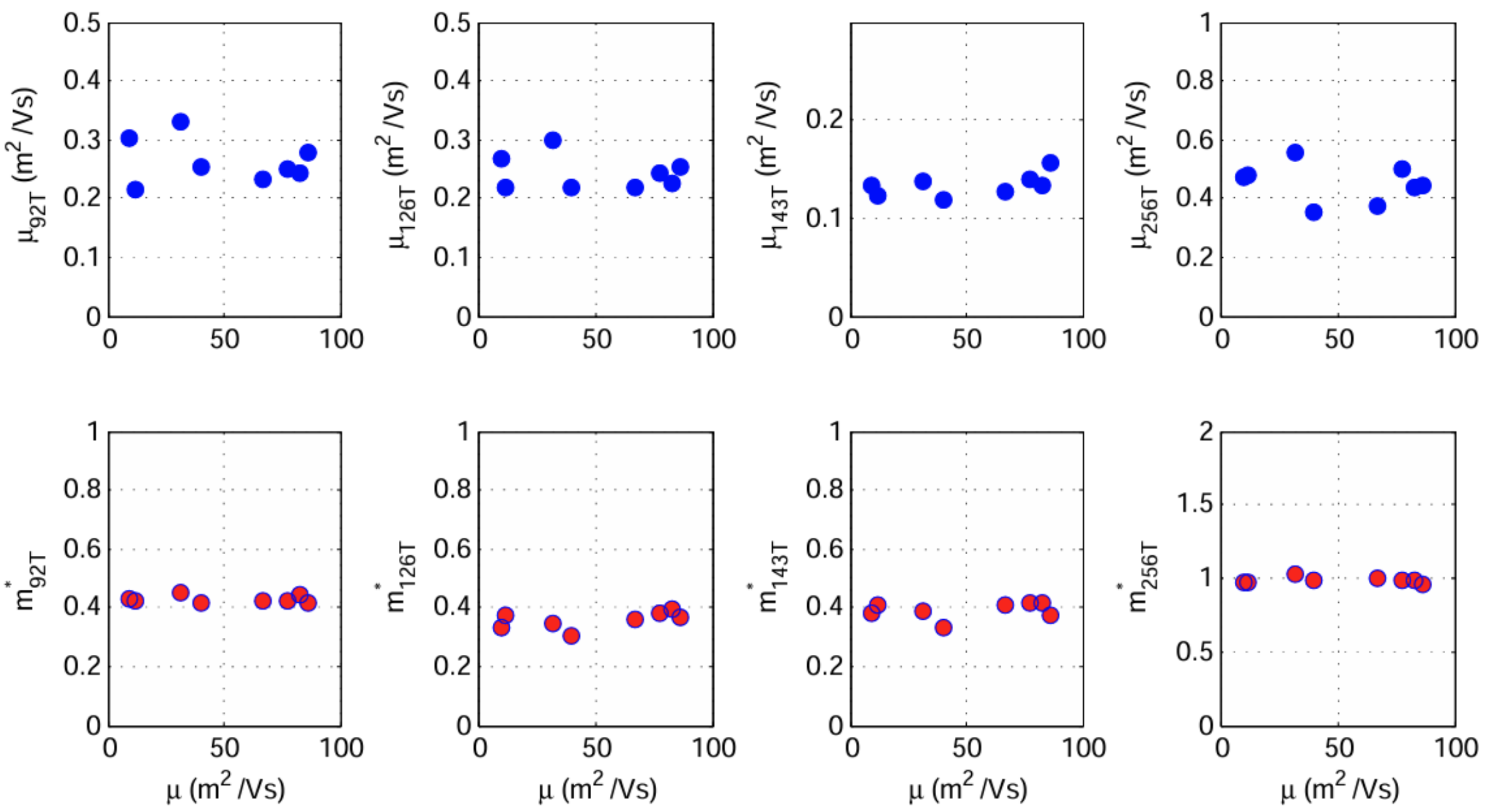

**Figure S7.5**. Quantum mobility versus transport mobility extracted from several $WTe_2$ crystals at all four frequencies.

The fact that only the transport mobility increases while the quantum mobility stays relatively unchanged hints at the nature of the defects. Transport mobility is sensitive to large-angle scattering that more abruptly changes carrier momentum and energy and is more likely with high density atomic defects such as oxygen substitution. On the other hand, quantum mobility is sensitive to scattering at all angles. Improving the crystal quality reduces the large-angle scattering events, but some more distant defects such as charged defect, as well as interactions present in the semimetal system such as electron-hole interaction may still cause decoherence of the Landau level, which appears as reduced quantum mobility in the system. We note this is not the case for the monolayer in Fig. 2, as there is only one dominant carrier after electrostatic doping.